\documentclass[11pt]{JHEP3}

\newcommand{\Sl}{\sum\limits}

\newcommand{\bea}{\begin{eqnarray}}
\newcommand{\eea}{\end{eqnarray}}
\newcommand{\bean}{\begin{eqnarray*}}
\newcommand{\eean}{\end{eqnarray*}}
\newcommand{\nn}{\nonumber \\}

\def\W #1{\widetilde{#1}}
\def\WH #1{\widehat{#1}}

\def\a{{\alpha}}

\def\b{{\beta}}

\def\Label#1{\label{#1}%
  \smash{\hbox to0pt{\raise1ex\hbox{\tiny[#1]}\hss}}}

\usepackage{subfigure,axodraw, amsmath,graphicx}
\usepackage{amsfonts}
\usepackage{amssymb}

\title{BCJ Relation of Color Scalar Theory and KLT Relation of Gauge Theory }
\author{Yi-Jian Du${}^{a}$, Bo Feng${}^{abc}$, Chih-Hao Fu${}^{b}$  \\
$^a$\small Zhejiang Institute of Modern Physics, Zhejiang
University, Hangzhou, 310027, P. R. China\\$^b$\small Center of
Mathematical Science, Zhejiang University, Hangzhou, China \\
$^c$\small Kavli Institute for Theoretical Physics China, CAS,
Beijing 100190, China\\~\\ E-mails: \email{yjdu@zju.edu.cn},
\email{b.feng@cms.zju.edu.cn}, \email{zhihao.fu@cms.zju.edu.cn}
 }

\abstract{ We present a field theoretical proof of the conjectured
KLT relation which states that the full tree-level scattering
amplitude of gluons can be written as a product of color-ordered
amplitude of gluons and color-ordered amplitude of scalars with only
cubic vertex. To give a  proof we establish the KK relation and BCJ
relation  of color-ordered scalar amplitude using BCFW recursion
relation with nonzero boundary contributions. As a byproduct, an
off-shell version of fundamental BCJ relation is proved, which plays
an important role in our work.

}

\begin{document}
\maketitle
\section{Introduction}

Despite the apparently straightforward algorithm laid down by
perturbation field theory, computations of scattering amplitude in
Yang-Mills theory has been known to be a formidable task even at
tree level due to the enormous amount of Feynman graphs consist in
the amplitude. In \cite{Witten:2003nn} a remarkable breakthrough was
made by Witten from string theory inspired observations, which has
lead to series of development ever since. For a review on these
progress see, for example, \cite{Cachazo:2005ga}. In particular, an
on-shell recursion relation devised to cope with tree amplitudes was
derived by Britto, Cachazo, Feng and Witten (BCFW)
\cite{Britto:2004ap, Britto:2005fq} based on analytic properties of
the S-matrix. Applications of BCFW recursion technique have recently
been extended to loop-levels by various groups (see for example
\cite{Boels:2010nw}).  A generalization which will be useful for our
paper is the BCFW on-shell recursion relation with nonzero boundary
contributions discussed in  \cite{BCFW:boundary}.

Incidentally, a perhaps even greater challenge was found in the
attempt of computing gravity amplitudes. The lagrangian of Einstein
gravity produces infinitely many vertices under perturbation, making
a direct computation exhausting all Feynman graphs practically
impossible. Parallel to the development in Yang-Mills theory, an
unexpected shortcut was found from the string dual to the theory. By
taking field theory limit a set of relations was found by Kawai,
Lewellen and Tye (KLT) which expresses gravity amplitudes as
``squares'' of the seemingly unrelated color-ordered Yang-Mills
amplitudes \cite{KLT}. While in string theory the proof of KLT
follows from monodromy \cite{KLT, BjerrumBohr:2010hn}, the proof of
KLT relations can as well be done from a purely field theoretical
viewpoint using BCFW recursion relation \cite{BjerrumBohr:2010ta,
BjerrumBohr:2010zb, BjerrumBohr:2010yc, Feng:2010br}. In the proof,
another set of relations discovered by Bern, Carrasco and Johansson
(BCJ) \cite{Bern:2008qj} played an essential role. As a byproduct of
the proof of KLT relation, a new $(n-2)!$ symmetric KLT relation was
founded in \cite{BjerrumBohr:2010ta} and will be used in this paper.

The BCJ relation displays merely one of the many interesting
properties of color-ordered amplitudes in gauge theory. BCJ relation
and other properties such as color-order reversed relation,
$U(1)$-decoupling relation and
 Kleiss-Kuijf (KK) relations \cite{Kleiss:1988ne}
  have been investigated from the point of view of string
theory in \cite{BjerrumBohr:2009rd,Stieberger:2009hq} (see also some
related result in \cite{Mafra:2011kj} from the pure spinor string
theory) as well as from field theory in \cite{DelDuca:1999rs,
Feng:2010my, Jia:2010nz, Tye:2010kg, Chen:2011jx, Monteiro:2011pc}.
The dual form of BCJ relation to loop level has also been discussed
in \cite{Bern:2010ue, Bern:2010yg,Bern:2011ia}.

Although the KK, BCJ and KLT relations were initially constructed to
describe properties of gauge theory and gravity (pure and
supersymmetric), these relations were found to function well beyond
their original designs. For example,  KK relations were proven to
hold for  gauge theory coupled to gravitons
\cite{Chen:2010ct} at leading order %(although the  BCJ relation does
%not apply to this theory)
and BCJ relations were shown to hold for tree
amplitudes with gauge field coupled to two fermions or
two scalars\cite{Sondergaard:2009za}.
 A more nontrivial example is the
conjectured made by Bern, Freitas and Wong (BFW)
\cite{Bern:1999bx}\footnote{Recently, KK and BCJ like relations for
the remaining rational function of one-loop gluon amplitudes have
also been investigated in \cite{BjerrumBohr:2011xe}.}
 that the full tree-level amplitude of gauge
theory can be expressed in the form of a KLT relation as products
of two amplitudes, where one of the amplitudes
is the color-ordered gauge amplitude and the other one is the amplitude of
a scalar theory that contains only cubic vertex described by a totally anti-symmetric
coupling constant $f^{abc}$.

The conjecture of BFW is simple but  not so easily  derived.
An initial reaction may be to use the supersymmetric KLT relation

\begin{equation}
\mathcal{M}^{\mathcal{N}=8}=\sum_{\alpha,\beta}
A^{\mathcal{N}=4}(\alpha)\mathcal{S}[\alpha|\beta]
\tilde{A}^{\mathcal{N}=4}(\beta),\label{eq:N8KLT}\end{equation}
since on one side of the equation, the full supermultiplet of ${\cal N}=8$ SUGRA
theory labeled by eight Grassmann variables $\eta^i, i=1,...,8$
contains both graviton and gluon, while one the other side of
the equation the full
supermultiplet of ${\cal N}=4$ SYM theory labeld by four Grassmann
variables ($\eta^i, i=1,...,4$ for $A$ and $\eta^i, i=5,...,8$ for
$\tilde{A}$) contains both gluon and scalar.
  Specifically, we obtain expressions of amplitudes of the desired
particle types by reading off the corresponding Grassmann
variable $\eta^{i_{1}}...\eta^{i_{r}}$ expansion coefficients with helicity of
the particle prescribed by $h=\mathcal{N}/4-r/2$. For the purpose of discussion
here, let us
associate $\eta^{1}\eta^{2}$ for gluons with positive helicity and
$\prod_{j=3}^{8}\eta^{j}$ for gluons with negative helicity on the
left hand side of (\ref{eq:N8KLT}). To keep the equation
consistent, on the right
hand side of (\ref{eq:N8KLT}) a contribution of $\eta^{1}\eta^{2}$
or $\eta^{3}\eta^{4}$ must come from amplitude $A^{\mathcal{N}=4}$ ,
whereas a contribution of zero power of $\eta$ or of
$\eta^{5}\eta^{6}\eta^{7}\eta^{8}$ must come from amplitude
$\tilde{A}^{\mathcal{N}=4}$, which means that $A^{\mathcal{N}=4}$
and $\tilde{A}^{\mathcal{N}=4}$ must be amplitudes of scalars and
gluons respectively.

Up to this step,  particle contents do match up those in
the  ${\cal N}=8$ KLT
relation. However, when  checking  the scalar part from
$A^{\mathcal{N}=4}$ carefully, we see immediately that it is not the pure
scalar theory with  cubic  vertex interaction. As a matter of fact
in ${\cal N}=4$ SYM theory a cubic vertex that links three
scalars is forbidden by charge conservation. Thus despite the many similarities
shared between KLT relation conjectured  by BFW  and the KLT
relations of the ${\cal N}=8$ theory,  the later can not be used to
derive former.

Although ${\cal N}=8$ KLT relation does not provide an explanation to
the conjecture, in principle, the
conjectured KLT relation can be derived from the field theory limit
of heterotic string theory as demonstrated in \cite{Tye:2010dd} with
careful calculations of cocycle part. So far we have not seen a
complete derivation along this line and it would be interesting to do
so rigorously.

In this paper we prove the KLT relation conjectured by BFW
\cite{Bern:1999bx}  from  field theory viewpoint. We
will follow similar steps in \cite{BjerrumBohr:2010ta,
BjerrumBohr:2010zb, BjerrumBohr:2010yc, Feng:2010br}. To be able to
do so first we take special cares on properties of scalar amplitudes.

The structure of this paper is organized as follows: In section
\ref{sec:BCFW}, we write down the BCFW recursion relation for the
scalar theory where the boundary contribution is nonzero. The nonzero
boundary contribution will bring many new features in our later
discussion. In section \ref{color-sec} we prove the color order reversed
identity and $U(1)$-decoupling identity  using  BCFW
recursion relation. Sections \ref{sec:KK} and \ref{polepart} are
devoted to proofs of KK and BCJ relations, where an off-shell
fundamental BCJ relation is established. Having  all the
preparations above, we finally prove the conjectured KLT relation in
section \ref{sec:KLT}. There we provide two proofs: one uses the
BCFW recursion and the other one,  a direct proof based on  the
off-shell fundamental BCJ relation. A conclusion and
discussions are given in the last section. Finally a generalized
U(1)-decoupling relation which explains the vanishing of boundary
contributions in KK relations is supplied in the Appendix.

%%%%%%%%%%%%%%%%%%%%%%%%%%%%%%%%%%%%%
%%%%%%%%%%%%%%%%%%%%%%%%%%%%%%%%%%%%%
%%%    gravity KLT relations
%%%
%%%%%%%%%%%%%%%%%%%%%%%%%%%%%%%%%%%%%%%
%%%%%%%%%%%%%%%%%%%%%%%%%%%%%%%%%%%%%%%

%%%%%%%%%%%%%%%%%%%%%%%%%%%%%%%%%%%%%%
%%%%%%%%%%%%%%%%%%%%%%%%%%%%%%%%%%%%%%
%%%
%%%  BCFW
%%%
%%%%%%%%%%%%%%%%%%%%%%%%%%%%%%%%%%%%%%
%%%%%%%%%%%%%%%%%%%%%%%%%%%%%%%%%%%%%%

\section{BCFW recursion relation and Boundary contribution in $\phi^{3}$ theory}
\label{sec:BCFW}

In the color-KLT relations proposed in \cite{Bern:1999bx},
color-dressed pure Yang-Mills amplitudes are expressed as a sum over
color-part scalar amplitudes multiplied by color-ordered amplitudes
of gluon. The scalar theory considered here contains only a
three-point vertex defined by the totally antisymmetric structure
constant $f^{abc}$ which satisfies Jacobi identity (we will call it
the "color-ordered scalar theory" from now on)
\begin{equation}
\sum_{b}(f^{a_{1}a_{2}b}f^{ba_{3}a_{4}}+f^{a_{1}a_{4}b}f^{ba_{2}a_{3}}+f^{a_{1}a_{3}b}f^{ba_{4}a_{2}})=0.
~~\label{Jacobi}\end{equation}
We will see that antisymmetry provides enough condition
for proving the relations to be discussed in sections \ref{color-sec}
and \ref{sec:KK}, but for  BCJ relation, the Jacobi-identity (\ref{Jacobi})
is crucial. The Jacobi-identity (\ref{Jacobi}), when expressed in terms
of three-point amplitudes, reads
\begin{equation}
A_{3}(1,2,-P_{12})A_{3}(-P_{34},3,4)+A_{3}(1,4,-P_{14})A_{3}(-P_{23},2,3)+A_{3}(1,3,-P_{13})
A_{3}(-P_{42},4,2)=0.~\label{Jacobi-in-A}\end{equation}
One thing worth  mentioning about equation (\ref{Jacobi-in-A}) is
that for the color-ordered scalar theory which contains only cubic vertex, the
three-point amplitude is the same for both on-shell and off-shell
momenta. This natural relation between off-shell momenta and
on-shell momenta for scalar field plays a very important role in many
of the properties to be discussed later. One reason for this simplicity is that
the wave function of scalar in momentum space is simply a number, $1$,
which does not depend on momentum.

The BCFW recursion relation of an amplitude is derived from applying
Cauchy's theorem to the complex integral
\begin{equation}
\int\frac{dz}{z}\, A(z)=A(0)+\sum_{poles\,
z_{i}}A(z_{i})=B,\label{eq:BCFWboundary}\end{equation}
where in the integrand a shifted amplitude $A(z)$ is given as an
analytic continuation defined by shifting a pair of the legs $(i,\,
j)$ by a light-like 4-vector, $\hat{p}_{i}=p_{i}+z\, q$,
$\hat{p_{j}}=p_{j}-zq$ with $q^2=q\cdot p_i=q\cdot p_j=0$.
Generically the integral (\ref{eq:BCFWboundary}) can have boundary
contribution $B$ if $A(z)$ does not vanish at infinity and the
unshifted amplitude therefore equals boundary contribution minus the
sum over residues taken at finite $z_{i}$, which assume the forms as
cut amplitudes $A_{L}(z_{i})\frac{1}{P^{2}}A_{R}(z_{i})$
\cite{Britto:2004ap, Britto:2005fq}. From the structure of Feynman
diagrams we see that when legs $(i,\, j)$ are not adjacent there
will be at least one propagator carrying a factor of ${1\over z}$
bridging between the two shifted legs so that $A(z)$ vanishes at
infinity. However boundary terms need to be taken into account when
$(i,\, j)$ are adjacent \cite{BCFW:boundary}. For example when the
two shifted legs are $p_{1}$ and $p_{n}$ we have
\bea
B=A_{L}(n,1,-P)\frac{1}{s_{n1}}A_{R}(P,2,...,n-1).~~\label{Boundary}\eea
 In the
expression  (\ref{Boundary}) above, the leg $P$ in $A(P,...)$ is
off-shell in general. However, as we have mentioned before, being
different from vectors and spinors, the scalar wave function does
not depends on the momentum and there is a natural analytic
continuation from on-shell amplitudes to off-shell amplitude, thus
boundary (\ref{Boundary}) with off-shell momentum $P$ does not
bother us much.

%%The numerator is only constructed by the color factors. When we
%%consider the color-order reversed relation, $U(1)$-decoupling
%%identity and KK relation in the following sections, the starting
%%point is only the antisymmetry of the structure constant, thus in
%%the recursive proof, we can also use lower-point relations for
%%$A(P,...)$. In the proof of the BCJ relation, things become more
%%complicated, because we have nontrivial kinematic factors before the
%%amplitudes.

As a warm up let us perform a consistent check by calculating the
same four-point scalar amplitude $A_{4}(1234)$ from all three possible
pairs of shiftings. It is easy to see that boundary contribution arise
when we shift $(1,2)$, and
\begin{equation}
A_{4}^{(12)}(1,2,3,4)=\frac{\sum_{b}f^{a_{4}a_{1}b}f^{ba_{2}a_{3}}}{s_{41}}+\frac{\sum_{b}f^{a_{1}a_{2}b}f^{ba_{3}a_{4}}}{s_{12}},\label{4-point-12}\end{equation}
where the first term in (\ref{4-point-12}) comes from cutting
propagator $1/S_{41}$ and the second arises as a boundary term.
Exactly the same formula is obtained from shifting $(4,1)$, whereas
in this case the roles played by the two terms in (\ref{4-point-12})
are swapped. The $(1,3)$-shifting is a bit different since the legs
are not adjacent. In this case both terms are given as cuts and we
obtain the same result for $A_{4}^{(13)}$.

We note that in the four-point example here the consistency among
three different pairs of shiftings does not lay further constraints
on structure of the coupling $f^{abc}$. The same is not true with
gauge theory. In \cite{Benincasa:2007xk}, Cachazo and Benincasa have
shown that the the same consistency at four-point requires
nontrivial Jacobi identity\footnote{See
\cite{Schuster:2008nh,He:2008nj} for generalization to arbitrary
$n$-gluons.}. Also note that highly nontrivial identities derived from
consistency requirement imposed by taking different choices of pairs of shiftings and  the
cancelation of spurious poles, have played important role in recent
development of scattering amplitudes of ${\cal N}=4$ theory  as
emphasized by \cite{Hodges:2009hk,ArkaniHamed:2009dn}.

%%%%%%%%%%%%%%%%%%%%%%%%%%%%%%%%%%%%%
%%%%%%%%%%%%%%%%%%%%%%%%%%%%%%%%%%%%%
%%%  color-order reversed, U(1)-decoupling, KK relations
%%%
%%%%%%%%%%%%%%%%%%%%%%%%%%%%%%%%%%%%%%%
%%%%%%%%%%%%%%%%%%%%%%%%%%%%%%%%%%%%%%%

\section{Color-order reversed relation and $U(1)$-decoupling relation}
\label{color-sec}
%%%%%%%%%%%%%%%%%%%%%%%%%%%%%%
Although for gauge theory color-order reversed relation and
$U(1)$-decoupling relation are somewhat trivial, it is not so for
color-ordered scalar theory, especially there is no natural
"$U(1)$-decoupling" analog.

First let us consider the color-order reverse relation, which is
given by
 \bea A(1,2,...,n)=(-1)^nA(n,n-1,...,1).~~\label{eq:c-o-r}
\eea
We prove the relation by induction. The staring point is the three-point
amplitude given by
 \bea
 A(1,2,3)=f^{a_1a_2a_3}=-f^{a_3 a_2a_1}=-A(3,2,1)~,
 \eea
We see that equation (\ref{eq:c-o-r}) is
clearly satisfied by the totally antisymmetric
 coupling constant $f^{abc}$. For general $n$, we  expand the
amplitude  by BCFW recursion relation with $(1,n)$-shifting as
\bea A(1,2,...,n)& = & \sum_{i=2}^{n-2}\sum_a A(\WH 1,2,..,i, -\WH
P_{1i}^a) {1\over s_{1i}}A(\WH P_{1i}^a, i+1,..,\WH n)\nn & &
+\sum_a A(n,1,-P_{1,n}^a)\frac{1}{s_{1,n}}A(P_{1,n}^a,2,...,n-1).
\eea
where the second line came from boundary contribution and
 a color $a$ is understood to be summed over. Applying the color-order reversed
relation to every sub-amplitude in above expression and recognizing
that it is nothing but the BCFW expansion of another amplitude,  we get
immediately the color order reversed relation
\bea A(1,2,...,n)&=& (-)^{n+2} \sum_{i=2}^{n-2}\sum_a  A(\WH n,
n-1,..., i+1, \WH P_{1i}^a)  {1\over s_{1i}}A(\WH P_{1i}^a,
i,...,\WH 1))\nn & & +\sum_a  A(n-1,..,2,P_{1,n}^a) {1\over s_{1,n}}
A(-P_{1,n}^a,1,n )\nn &=&(-1)^n A(n,n-1,...,1), \eea
where we see that only the antisymmetric property of three-point
amplitude and the applicability of   the BCFW recursion relation
with boundary were needed in deriving this result.

Next let us we move on to the  $U(1)$-decoupling identity given by
\bea
A(1,2,3,...,n)+A(1,3,2,...,n)+...+A(1,3,4,...,n,2)=0.~~\label{eq:u1-d}
\eea
The simplest $U(1)$-decoupling identity is for $n=3$, which is
nothing but the color-order reversed identity. The next simplest
case for four-point amplitude can be checked directly using
 the explicit result given in (\ref{4-point-12})
\bean I & = & A_4(1,2,3,4)+A_4(1,3,2,4)+ A_4(1,3,4,2)\nn & = &
\left[ {\sum_b f^{a_4 a_1 b} f^{b a_2 a_3}\over s_{41}} + {\sum_b
f^{a_1 a_2 b} f^{b a_3 a_4}\over s_{12}}\right] + \left[ {\sum_b
f^{a_4 a_1 b} f^{b a_3 a_2}\over s_{41}} + {\sum_b f^{a_1 a_3 b}
f^{b a_2 a_4}\over s_{13}}\right]\nn
& & + \left[ {\sum_b f^{a_2 a_1 b} f^{b a_3 a_4}\over s_{12}} +
{\sum_b f^{a_1 a_3 b} f^{b a_4 a_2}\over s_{13}}\right]\nn
& = & {\sum_b f^{a_4 a_1 b} f^{b a_2 a_3}+\sum_b f^{a_4 a_1 b} f^{b
a_3 a_2}\over s_{41}} + {\sum_b f^{a_1 a_2 b} f^{b a_3 a_4}+\sum_b
f^{a_2 a_1 b} f^{b a_3 a_4}\over s_{12}}\nn
& & + {\sum_b f^{a_1 a_3 b} f^{b a_4 a_2}+\sum_b f^{a_1 a_3 b} f^{b
a_2 a_4}\over s_{13}}\nn
& = & 0,~\eean
where again, only the totally anti-symmetric property of coupling
constant $f^{abc}$ is used.

For general $n$ we derive an inductive proof via  $(1,2)$-shifting.
For $3\leq k\leq n-1$, the BCFW-expansion of a general $n$-point amplitude is
(We drop the summation over color index $a$
for simplicity.)
\bea A(1,3,..,k,2,k+1,...,n) & = &\sum_{n\geq j\geq k+1} \sum_{3\leq
i\leq k}A_L (j,j+1,..,n,1,3,..,i,P_{ji}) {1\over P_{ji}^2}
A_R(-P_{ji}, i+1,..,k,2,...,j-1)\nn
& & + \sum_{n\geq j\geq k+1} A_L (j,j+1,..,n,1,P_{j1}) {1\over
P_{j1}^2} A_R(-P_{j1}, 3,..,k,2,...,j-1)\nn
& & +  \sum_{3\leq i\leq k}A_L (1,3,..,i,P_{1i}) {1\over P_{1i}^2}
A_R(-P_{1i}, i+1,..,k,2,...,n),~~~\label{U1-n-gen-k}\eea
 where we have isolated  in the last two lines terms
that contain a cut adjacent to leg $1$.
There are also two special amplitudes that produce
nonzero  boundary contributions under  BCFW-expansion:
\bea A(1,2,3,...,n) & = & A_L(1,2,P){1\over s_{12}} A_R(-P,3,4,..,n)
\nn & & +  \sum_{n\geq j\geq 4} A_L (j,j+1,..,n,1,P_{j1}) {1\over
P_{j1}^2} A_R(-P_{j1}, 2,3,...,j-1)~~~\label{U1-n-gen-b1}\eea
and
\bea A(1,3,4,..,n,2) & = & A_L(2,1,P){1\over s_{12}}A_R(-P,3,4,..,n)
\nn & & +\sum_{3\leq i\leq n-1}A_L (1,3,..,i,P_{1i}) {1\over
P_{1i}^2} A_R(-P_{1i}, i+1,..,n,2)~~~\label{U1-n-gen-b2} \eea
Having written down  BCFW-expansion for every amplitudes,  we sum them up and
collect various contributions for given cut momentum. There are two
contributions for the cut momentum ${ P_{12}}$ coming from
(\ref{U1-n-gen-b1}) and (\ref{U1-n-gen-b2}) and we have
\bean [A_L(1,2,P)+A_L(2,1,P)]{1\over s_{12}} A_R(-P,3,4,..,n)=0\eean
where the anti-symmetry property of three-point amplitudes was
used. For the cut momentum $P_{1i}$ with $3\leq i\leq n-1$, both
(\ref{U1-n-gen-k}) and (\ref{U1-n-gen-b2}) have contributions as
\bean A_L (1,3,..,i,P_{1i}) {1\over P_{1i}^2}\left\{ A_R(-P_{1i},
i+1,..,n,2)+\sum_{i\leq k\leq n-1} A_R(-P_{1i},
i+1,..,k,2,...,n)\right\}=0~,\eean
where we have used  $U(1)$-decoupling identity for the part inside
the big curly bracket by induction. For the cut momentum $P_{j1}$
both (\ref{U1-n-gen-k}) and (\ref{U1-n-gen-b1}) have contributions
as
\bean A_L (j,j+1,..,n,1,P_{j1}){1\over P_{j1}^2}\left\{ A_R(-P_{j1},
2,3,...,j-1)+ \sum_{3\leq k\leq j-1} A_R(-P_{j1},
3,..,k,2,...,j-1)\right\}=0\eean
where again by the induction,  $U(1)$-decoupling identity has been
applied to the part inside the big curly bracket. Finally for the cut
momentum $P_{ji}$ with $n\geq j\geq 4$
 $3\leq i\leq n-1$ and $j-i\geq 2$, only (\ref{U1-n-gen-k}) gives contribution
 with $i\leq k\leq k-1$, thus we have
\bean A_L (j,j+1,..,n,1,3,..,i,P_{ji}) {1\over P_{ji}^2} \sum_{i\leq
k \leq j-1} A_R(-P_{ji}, i+1,..,k,2,...,j-1)=0\eean
where  the induction is applied to the part  $\sum_k A_R$.

Before concluding this section, we would like to emphasize again
that both the color-order reserved identity and $U(1)$-decoupling
identity used only the anti-symmetric property of coupling constant
$f^{abc}$ and the BCFW recursion relation with boundary
contribution. Especially the Jacobi identity (\ref{Jacobi}) is not
needed for these two properties. Also, when we take the external
momentum from on-shell to off-shell, these two identities hold too.
These facts will be used in our proof of BCJ relation.

%%%%%%%%%%%%%%%%%%%%%%%%
\section{The KK relation for color-ordered scalar theory}\label{sec:KK}
%%%%%%%%%%%%%%%%%%%%%%%%%%%

In this section, we will prove that for the color-ordered scalar
theory, there is a similar KK-relation found originally in
\cite{Kleiss:1988ne} for gauge theory:
%scalar KK relation by BCFW
%recursion relation with nontrivial boundary contribution. The KK
%relation \cite{Kleiss:1988ne}is
%
\bea
A_n(\beta_1,...,\beta_r,1,\alpha_1,...,\alpha_s,n)=(-1)^{r}\Sl_{\{\sigma\}\in
P(O\{\alpha\}\cup O\{\beta\}^T)}A_n(1,\{\sigma\},n),~~\label{KK}
 \eea
where the sum is over all permutations keeping relative ordering
inside the set $\a$ and the set $\b^T$ (where the $T$ means the
reversing ordering of the set $\b$), but allowing all relative
ordering between the set $\a$ and $\b$. When there is no element in
the $\{\a\}$, (\ref{KK}) reduces to the color-order reversed
relation\eqref{eq:c-o-r}, while  when there is only one element in
the set $\{\beta\}$, (\ref{KK}) reduces to the $U(1)$-decoupling
identity\eqref{eq:u1-d}. Thus the color-order reversed relation and
the $U(1)$-decoupling identity are just the special cases of KK
relation. If KK-relation is true, above two identities are also
true, but the reverse is not guaranteed and it is possible that both
identities are true, but the KK-relation is not true\footnote{We do
not have any example yet and it will be interesting to see if there
is an example.}.  Since for $n\leq 5$, there is no true KK-relation
not covered by the color-order reversed relation and the
$U(1)$-decoupling identity, the starting point of induction proof is
checked.

Now we give the general proof of the KK relation by BCFW recursion
relation with shifting pair $(1,n)$. The idea will be same as the
one used in \cite{Feng:2010my}, but there is one major difference: at
the left hand side of (\ref{KK}), there is no boundary contribution
since we have assumed $r\geq 2$, while at the right hand side, each
amplitude will give a boundary contribution. Since from the proof
given in \cite{Feng:2010my}, we see that the pole parts of the left
hand side and the right hand side match up, thus for (\ref{KK}) to
be true, the sum of all boundary contributions must be zero.

Having the above general picture, we repeat some steps given in
\cite{Feng:2010my} for self-completion. Using the BCFW recursion
relation for scalar field, the amplitude
$A_n(\beta_1,...,\beta_r,1,\alpha_1,...,\alpha_s,n)$ can be written
as (it is worth to notice that there is no boundary contribution)
\bea
A_n(\beta_1,...,\beta_r,1,\alpha_1,...,\alpha_s,n)=\sum\limits_{\text{All
splittings}} A_n(\{\beta_L\},\WH 1,\{\alpha_L\}|\{\alpha_R\},\WH
n,\{\beta_R\}), \eea
where we have used the short notation $A(\a|\b)\equiv A_L (\a, -\WH
P_\a){1\over P_\a^2} A_R(\WH P_\a, \b)$. For each splitting
$A(\a|\b)$ we can use the KK-relation for $A_L (\a, -\WH P_\a)$ and
$A_R(\WH P_\a, \b)$ by our induction, thus we have
 \bea && A_n(\beta_1,...,\beta_r,1,\alpha_1,...,\alpha_s,n)\nn
&=&(-1)^{n_{\beta_L}+n_{\beta_R}}\sum\limits_{\text{All
splittings}}\sum_{\{\sigma_L\in P(O\{\alpha_L\}\cup
O\{\beta_L\}^T)\}}\sum_{\{\sigma_R\in P(O\{\alpha_R\}\cup
O\{\beta_R\}^T)\}} A_n(\WH 1,\{\sigma_L\}|\{\sigma_R\},\WH n)\nn
&=&(-1)^{r}\sum\limits_{\{\sigma\}\in P(O\{\alpha\}\cup
O\{\beta^T\})}\sum_{\text{All splittings for  $\sigma$}}A_n(\WH
1,\{\sigma_L\}|\{\sigma_R\},\WH n). \eea
It is worth to notice that for the scalar theory with $1,n$ nearby,
there is boundary contribution for each amplitude, thus the last
line in above equation should be rewritten as
\bea\label{nptKKwithboundary}
 &&A_n(\beta_1,...,\beta_r,1,\alpha_1,...,\alpha_s,n)\nn
 &=&(-1)^{r}\sum\limits_{\{\sigma\}\in
P(O\{\alpha\}\cup O\{\beta^T\})}\left\{A_n(\WH 1,\{\sigma\},\WH
n)-A_3(n,1,-P_{1,n})\frac{1}{P_{1,n}^2}A_{n-1}(P_{1,n},\{\sigma\})\right\}.\nn
\eea
Thus for the KK-relation (\ref{KK}) to be true, we must show
\bea \sum\limits_{\{\sigma\}\in P(O\{\alpha\}\cup
O\{\beta^T\})}A_{n-1}(P_{1,n},\{\sigma\})=0~. \label{Two-KK}\eea
Equation (\ref{Two-KK}) is easy to show by using the KK-relation
with less number of particles. To see it, assuming the ordering of
set $\a$ is $\a=\{\a_1,..,\a_k\}$ and the set $\b^T$ is $\b^T=\{
\b_1,...,\b_m\}$, then in the sum $\sum_{P(O(\a)\bigcup O(\b^T))}$
all allowed orderings can be divided into two cases: the first case
is that the last element is $\a_k$ and the second case is that the
last element is $\b_m$. All terms of the first case can be sum up to
\bea I_1=(-)^m A(1, \{\a_1,...,\a_{k-1}\}, \a_k, \{ \b_m,
\b_{m-1},...,\b_1\})~. \eea
while  all terms of the second case can be sum up to
\bea I_2=(-)^{m-1} A(1, \{\a_1,...,\a_{k-1},\a_k\}, \b_m, \{
\b_{m-1},...,\b_1\})~.\eea
It is then easy to see that (\ref{Two-KK}) equals to
\bea I_1+ I_2=0~.\eea

The proof of (\ref{Two-KK}) used the KK-relation, thus it is also
true for gauge theory. In fact, as we will explain in the Appendix,
it has a very natural physical meaning, i.e.,  the generalized
$U(1)$-decoupling identity. Also, the momentum $P$ is off-shell in
(\ref{Two-KK}) does not matter to our proof, i.e., the KK-relation
is true with off-shell momentum with natural off-shell continuation.

%%%%%%%%%%%%%%%%%%%%%%%%%%%%%%%%%%%
%%%%%%%%%%%%%%%%%%%%%%%%%%%%%%%
%%%
%%%   BCJ
%%%
%%%%%%%%%%%%%%%%%%%%%%%%%%%%%%%%%%%
%%%%%%%%%%%%%%%%%%%%%%%%%%%%%%%%%%%
%%%%%%%%%%%%%%%%%%%%%%%%%%%%%%%%%%%

%%%%%%%%%%%%%%%%%%%%%%
\section{Fundamental BCJ relations of color-dressed scalar amplitudes}
\label{polepart}
%%%%%%%%%%%%%%%%%

In this section we  show that   the BCJ relation, which was found
by Bern, Carrasco and Johansson in \cite{Bern:2008qj}  for gauge
theory,   also applies to the color-ordered scalar theory provided
that Jacobi identity (\ref{Jacobi}) is satisfied by the coupling constant
$f^{abc}$.

To see the reason for introducing Jacobi identity (\ref{Jacobi}),  let us
consider the BCJ relation for four-point amplitudes
\begin{equation}
s_{21}\tilde{A}(1234)+(s_{21}+s_{23})\tilde{A}(1324)=0,
~~\label{4ptbcj}
\end{equation}
We decorate scalar amplitudes with ``tildes'' so that they are not
to be confused
with  gluon amplitudes when  we discuss KLT relations  in later
sections. Inserting explicit expressions of  four-point amplitudes, and we
have
\bea s_{12} \tilde{A}(1,2,3,4) & = & s_{12}\left[ {\sum_b f^{a_4 a_1 b} f^{b
a_2 a_3}\over s_{41}} + {\sum_b f^{a_1 a_2 b} f^{b a_3 a_4}\over
s_{12}}\right],\nn
s_{13} \tilde{A}(1,3,2,4) & = &  s_{13} \left[ {\sum_b f^{a_4 a_1 b}
f^{b a_3 a_2}\over s_{41}} + {\sum_b f^{a_1 a_3 b} f^{b a_2
a_4}\over s_{13}}\right],\eea
The left hand side of  equation (\ref{4ptbcj})   can therefore be written as
\bea  & & s_{12} \tilde{A}(1,2,3,4)- s_{13} \tilde{A}(1,3,2,4) \nn
& = &{s_{12}\sum_b f^{a_4 a_1 b} f^{b a_2 a_3}-s_{13}\sum_b f^{a_4
a_1 b} f^{b a_3 a_2}\over s_{41}}+ \sum_b f^{a_1 a_2 b} f^{b a_3
a_4}-\sum_b f^{a_1 a_3 b} f^{b a_2 a_4} ,\eea
where we have used conservation of momentum to rewrite
$(s_{12}+s_{32})=\,-s_{13}$. To carry out rest of the calculation we
note that Jacobi identity is essential in combining the last two
terms, giving
\bea & & s_{12} \tilde{A}(1,2,3,4)- s_{13} \tilde{A}(1,3,2,4) \nn
& = &{s_{12}\sum_b f^{a_4 a_1 b} f^{b a_2 a_3}-s_{13}\sum_b f^{a_4
a_1 b} f^{b a_3 a_2}\over s_{41}}- \sum_b f^{a_1 a_4 b} f^{b a_2
a_3}\nn
& = &{s_{12}\sum_b f^{a_4 a_1 b} f^{b a_2 a_3}-(s_{13}+s_{14})\sum_b
f^{a_4 a_1 b} f^{b a_3 a_2}\over s_{41}}\nn
& = & {s_{12}\sum_b f^{a_4 a_1 b} f^{b a_2 a_3}+s_{12}\sum_b f^{a_4
a_1 b} f^{b a_3 a_2}\over s_{41}}\nn
& = & 0~.\eea

Since KK-relation is satisfied by the color-ordered scalar theory, to show
the general BCJ relation is also satisfied we only need to prove
the following fundamental
BCJ relation
\bea 0=I_{n}=s_{21} A(1,2,...,n)+\sum_{i=3}^{n-1}
(s_{21}+\sum_{t=3}^i s_{2t})
A(1,3,..,i,2,i+1,...,n-1,n).~~\label{BCJ-fund} \eea
To prove (\ref{BCJ-fund}) we deform $I_n$  by shifting the pair of legs $(1,n)$, yielding
\begin{equation}
I_{n}(z)=s_{2\hat{1}}\tilde{A}(\hat{1}23\dots\hat{n})+(s_{2\hat{1}}+s_{23})\tilde{A}(\hat{1}32\dots\hat{n})+\dots(s_{2\hat{1}}+\dots
s_{2,n-1})\tilde{A}(\hat{1}3\dots
n-1,\,\hat{n})\label{eq:ieqn}\end{equation}
and then consider the following contour integration
\begin{equation}
B_n\equiv \int\frac{dz}{z}\, I_{n}(z)=I_{n}(0)+\sum_{poles\,
z_{i}}I_{n}(z_{i})~\label{eq:intbcj}\end{equation}
The contour integration can be done by two different ways. The first
way is to integrate around the pole at infinity  and the
result is the boundary contribution $B_n$. The second way is to
integrate along large enough circle around point $z=0$ and we obtain the right
hand side of (\ref{eq:intbcj}), which includes two parts, i.e., the
$I_n$ part and the pole parts.

In the proof of BCJ relation for gauge theory,  the ${1\over z^2}$
behavior of gauge amplitudes with nonadjacent shifted legs is
crucial to derive $B_n=0$ in \cite{Feng:2010my}. Then by the
induction, we can show  the pole part is zero, from there we
conclude that $I_n=0$. For color-ordered scalar theory, it is again
easy to show the pole part to be zero by induction, thus to prove
$I_n=0$ we must show $B_n=0$. However, for the scalar theory, the
large $z$ behavior of amplitudes with non-nearby shifting pair is
${1\over z}$, thus many boundary terms are expected. The cancelation
of these boundary terms relies crucially on the Jacobi identity as
we will see shortly.

Having the above general picture, we will present our proof in two
steps: First we will show that the pole part is zero, and then we will show
that the boundary contribution is zero.

%%%%%%%%%%%%%%%%%%%%%%
\subsection{The pole part}
%%%%%%%%%%%%%%%%%%%%%%%%

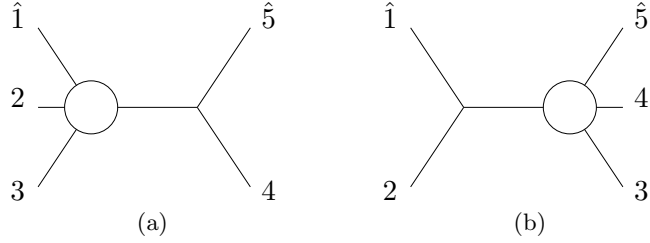
\begin{figure}[!h]
  \centering
  %%%%%%%%%%%%%%%%%%%%
  %%  a
  %%%%%%%%%%%%%%%%%%%%
    \subfigure[]{
% \fcolorbox{white}{white}{
  \begin{picture}(130,70) (-20,0)
    \SetWidth{0.375}

\Line(20,30)(0,0)
\Line(20,30)(0,30)
\Line(20,30)(0,60)

\Line(20,30)(60,30)

\Line(60,30)(80,0)
%\Line(60,30)(80,30)
\Line(60,30)(80,60)

    \BCirc(20,30){10}
%    \BCirc(60,30){10}

        \Text(-5,60)[br]{$\hat{1}$}
        \Text(-5,30)[br]{$2$}
        \Text(-5,-5)[br]{$3$}
        \Text(90,-5)[br]{$4$}
%        \Text(90,30)[br]{$4$}
        \Text(90,60)[br]{$\hat{5}$}
  \end{picture}\label{grapha}
}
  %%%%%%%%%%%%%%%%%%%%
  %%  b
  %%%%%%%%%%%%%%%%%%%%
    \subfigure[]{
% \fcolorbox{white}{white}{
  \begin{picture}(130,70) (-20,0)
    \SetWidth{0.375}

\Line(20,30)(0,0)
%\Line(20,30)(0,30)
\Line(20,30)(0,60)

\Line(20,30)(60,30)

\Line(60,30)(80,0)
\Line(60,30)(80,30)
\Line(60,30)(80,60)

%    \BCirc(20,30){10}
    \BCirc(60,30){10}

        \Text(-5,60)[br]{$\hat{1}$}
%        \Text(-5,30)[br]{$2$}
        \Text(-5,-5)[br]{$2$}
        \Text(90,-5)[br]{$3$}
        \Text(90,30)[br]{$4$}
        \Text(90,60)[br]{$\hat{5}$}
  \end{picture} \label{graphb}
}
\end{figure}

In this part of the discussion we  consider the contributions from pole part.
Before going to general $n$, let us check the example when $n=5$. The
pole part contribution from $\tilde{A}(\hat{1}234\hat{5})$ is given by
figures (\ref{grapha}) and (\ref{graphb}).

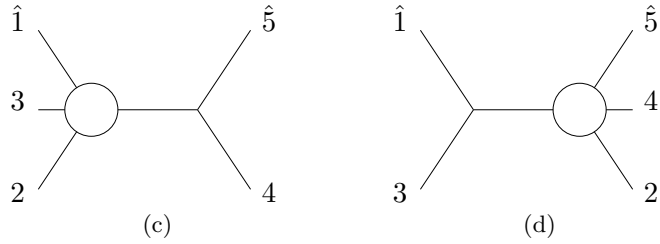
\begin{figure}[!h]
  \centering
  %%%%%%%%%%%%%%%%%%%%
  %%  c
  %%%%%%%%%%%%%%%%%%%%
    \subfigure[]{
% \fcolorbox{white}{white}{
  \begin{picture}(130,70) (-20,0)
    \SetWidth{0.375}

\Line(20,30)(0,0)
\Line(20,30)(0,30)
\Line(20,30)(0,60)

\Line(20,30)(60,30)

\Line(60,30)(80,0)
%\Line(60,30)(80,30)
\Line(60,30)(80,60)

    \BCirc(20,30){10}
%    \BCirc(60,30){10}

        \Text(-5,60)[br]{$\hat{1}$}
        \Text(-5,30)[br]{$3$}
        \Text(-5,-5)[br]{$2$}
        \Text(90,-5)[br]{$4$}
%        \Text(90,30)[br]{$4$}
        \Text(90,60)[br]{$\hat{5}$}
  \end{picture} \label{graphc}
}
  %%%%%%%%%%%%%%%%%%%%
  %%  d
  %%%%%%%%%%%%%%%%%%%%
    \subfigure[]{
% \fcolorbox{white}{white}{
  \begin{picture}(130,70) (-20,0)
    \SetWidth{0.375}

\Line(20,30)(0,0)
%\Line(20,30)(0,30)
\Line(20,30)(0,60)

\Line(20,30)(60,30)

\Line(60,30)(80,0)
\Line(60,30)(80,30)
\Line(60,30)(80,60)

%    \BCirc(20,30){10}
    \BCirc(60,30){10}

        \Text(-5,60)[br]{$\hat{1}$}
%        \Text(-5,30)[br]{$2$}
        \Text(-5,-5)[br]{$3$}
        \Text(90,-5)[br]{$2$}
        \Text(90,30)[br]{$4$}
        \Text(90,60)[br]{$\hat{5}$}
  \end{picture} \label{graphd}
}
\end{figure}

Similarly, the pole part from $\tilde{A}(\hat{1}324\hat{5})$ is
given by (\ref{graphc}) and (\ref{graphd})

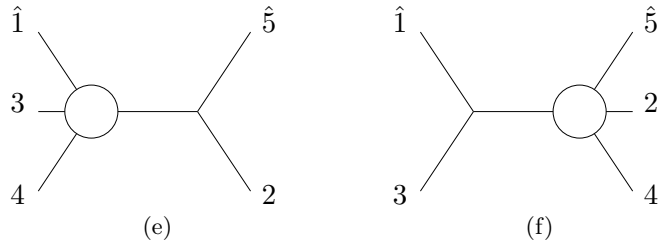
\begin{figure}[!h]
  \centering
    %%%%%%%%%%%%%%%%%%%%
  %%  e
  %%%%%%%%%%%%%%%%%%%%
    \subfigure[]{
% \fcolorbox{white}{white}{
  \begin{picture}(130,70) (-20,0)
    \SetWidth{0.375}

\Line(20,30)(0,0)
\Line(20,30)(0,30)
\Line(20,30)(0,60)

\Line(20,30)(60,30)

\Line(60,30)(80,0)
%\Line(60,30)(80,30)
\Line(60,30)(80,60)

    \BCirc(20,30){10}
%    \BCirc(60,30){10}

        \Text(-5,60)[br]{$\hat{1}$}
        \Text(-5,30)[br]{$3$}
        \Text(-5,-5)[br]{$4$}
        \Text(90,-5)[br]{$2$}
%        \Text(90,30)[br]{$4$}
        \Text(90,60)[br]{$\hat{5}$}
  \end{picture} \label{graphe}
}
  %%%%%%%%%%%%%%%%%%%%
  %%  f
  %%%%%%%%%%%%%%%%%%%%
    \subfigure[]{
% \fcolorbox{white}{white}{
  \begin{picture}(130,70) (-20,0)
    \SetWidth{0.375}

\Line(20,30)(0,0)
%\Line(20,30)(0,30)
\Line(20,30)(0,60)

\Line(20,30)(60,30)

\Line(60,30)(80,0)
\Line(60,30)(80,30)
\Line(60,30)(80,60)

%    \BCirc(20,30){10}
    \BCirc(60,30){10}

        \Text(-5,60)[br]{$\hat{1}$}
%        \Text(-5,30)[br]{$2$}
        \Text(-5,-5)[br]{$3$}
        \Text(90,-5)[br]{$4$}
        \Text(90,30)[br]{$2$}
        \Text(90,60)[br]{$\hat{5}$}
  \end{picture} \label{graphf}
}
\end{figure}

and finally the pole part from $\tilde{A}(\hat{1}342\hat{5})$ is
given by (\ref{graphe}) and (\ref{graphf}).

\vspace{1cm}

It is easy to see that a factor $\frac{1}{P_{45}}\,
\tilde{A}(\hat{5},4,-P_{4\hat{5}})$ is shared between graphs
\ref{grapha} and \ref{graphc}, thus the combination of these two
contributions  gives
\begin{equation}
s_{2\hat{1}}\tilde{A}(\hat{1},2,3,-P_{\hat{1}23})+(s_{2\hat{1}}+s_{23})\tilde{A}(\hat{1},3,2,-P_{\hat{1}23}),\end{equation}
which is zero using 4-point BCJ relation, as was verified
explicitly. Again using 4-point BCJ relation we see that the sum of
graphs \ref{graphd} and \ref{graphf} is zero after  using momentum
conservation to rewrite
$(s_{2\hat{1}}+s_{23})=-(s_{24}+s_{2\hat{5}})$ and
$(s_{2\hat{1}}+s_{23}+s_{24})=-s_{2\hat{5}}$. Finally the graphs
\ref{graphb} and \ref{graphe} vanish individually because the
shifted internal momenta become null at pole, i.e., from the BCJ relation
of 3-point amplitude, $s_{12}\tilde{A}(1,2,3)=0$.

\vspace{1cm}

Generalization of the 5-point example to n-point BCJ relation is
straightforward. Pole contributions from  (\ref{eq:ieqn})  fall into
one of following two categories: Those with leg $2$ attached to the
sub-amplitude on the left hand side of the BCFW-expansion and those
to the sub-amplitude on the right hand side of the BCFW-expansion.
Graphs with leg $2$ attached to the left hand side with split
momentum $P_{12...i}$ add up to give zero from $i+1$-point BCJ
relation,
\begin{figure}[!h]
  \centering
  %%%%%%%%%%%%%%%%%%%%
  %%  pole a
  %%%%%%%%%%%%%%%%%%%%
    \subfigure{
% \fcolorbox{white}{white}{
  \begin{picture}(140,80) (10,-10)
    \SetWidth{0.375}

\Line(20,30)(0,0)
\Line(20,30)(0,45)
\Line(20,30)(0,60)

\Line(20,30)(60,30)

\Line(60,30)(80,0)
%\Line(60,30)(80,30)
\Line(60,30)(80,60)
%      \Line(70,20)(90,10)
    \BCirc(20,30){8}
    \BCirc(60,30){8}

        \Text(-5,60)[br]{$\hat{1}$}
        \Text(-5,45)[br]{$2$}
        \Text(-5,-5)[br]{$i$}
                \Text(90,-10)[br]{$i+1$}
          \Text(-10,40)[br]{$s_{2\hat{1}}$}
                   \Text(0,20)[br]{$\vdots$}
                   \Text(85,30)[br]{$\vdots$}
%        \Text(40,35)[br]{$\frac{1}{z}$}
        \Text(90,60)[br]{$\hat{n}$}
                \Text(120,20)[br]{$+\dots$}
  \end{picture}
}
    \subfigure{
% \fcolorbox{white}{white}{
  \begin{picture}(100,80) (-50,-10)
    \SetWidth{0.375}

\Line(20,30)(0,0)
\Line(20,30)(0,15)
\Line(20,30)(0,60)

\Line(20,30)(60,30)

\Line(60,30)(80,0)
%\Line(60,30)(80,30)
\Line(60,30)(80,60)
%      \Line(70,20)(90,10)
    \BCirc(20,30){8}
    \BCirc(60,30){8}

        \Text(120,20)[br]{$=0$}
        \Text(-5,60)[br]{$\hat{1}$}
        \Text(-5,15)[br]{$i$}
        \Text(-5,-5)[br]{$2$}
                \Text(90,-10)[br]{$i+1$}
          \Text(-10,40)[br]{$(s_{2\hat{1}}+\dots+s_{2i}) $}
                   \Text(0,30)[br]{$\vdots$}
                   \Text(85,30)[br]{$\vdots$}
%        \Text(40,35)[br]{$\frac{1}{z}$}
        \Text(90,60)[br]{$\hat{n}$}
  \end{picture}
} \label{npta}
\end{figure}

Flipping kinematic factors using momentum conservation as was done in the
5-point example, we see that graphs with leg $2$ attached to the
right hand side with split momentum $P_{13...i}$ also add up to
zero.

\begin{figure}[!h]
  \centering
  %%%%%%%%%%%%%%%%%%%%
  %%  pole a
  %%%%%%%%%%%%%%%%%%%%
    \subfigure{
% \fcolorbox{white}{white}{
  \begin{picture}(140,80) (-20,-10)
    \SetWidth{0.375}

\Line(20,30)(0,0)
\Line(60,30)(80,15)
\Line(20,30)(0,60)

\Line(20,30)(60,30)

\Line(60,30)(80,0)
%\Line(60,30)(80,30)
\Line(60,30)(80,60)
%      \Line(70,20)(90,10)
    \BCirc(20,30){8}
    \BCirc(60,30){8}

        \Text(-5,60)[br]{$\hat{1}$}
        \Text(105,10)[br]{$i+1$}
        \Text(-5,-5)[br]{$i$}
                \Text(90,-10)[br]{$2$}
          \Text(-10,40)[br]{$-(s_{2,i+1}+\dots+s_{2\hat{n}})$}
                   \Text(0,30)[br]{$\vdots$}
                   \Text(85,30)[br]{$\vdots$}
%        \Text(40,35)[br]{$\frac{1}{z}$}
        \Text(90,60)[br]{$\hat{n}$}
                \Text(140,20)[br]{$+\dots$}
  \end{picture}
}
    \subfigure{
% \fcolorbox{white}{white}{
  \begin{picture}(100,80) (-30,-10)
    \SetWidth{0.375}

\Line(20,30)(0,0)
\Line(60,30)(80,45)
\Line(20,30)(0,60)

\Line(20,30)(60,30)

\Line(60,30)(80,0)
%\Line(60,30)(80,30)
\Line(60,30)(80,60)
%      \Line(70,20)(90,10)
    \BCirc(20,30){8}
    \BCirc(60,30){8}

        \Text(120,20)[br]{$=0$}
        \Text(-5,60)[br]{$\hat{1}$}
        \Text(90,45)[br]{$2$}
        \Text(-5,-5)[br]{$i$}
                \Text(90,-10)[br]{$i+1$}
          \Text(-10,40)[br]{$-s_{2\hat{n}} $}
                   \Text(0,30)[br]{$\vdots$}
                   \Text(85,20)[br]{$\vdots$}
%        \Text(40,35)[br]{$\frac{1}{z}$}
        \Text(90,60)[br]{$\hat{n}$}
  \end{picture} \label{nptb}
}
\end{figure}
Thus, using the BCJ relation of $\tilde{A}_m$ for $m\leq n$, we have
shown that the pole part of (\ref{eq:ieqn}) is zero.

%%%%%%%%%%%%%%%%%%%%%
\subsection{the generalized fundamental BCJ relation with one off-shell momentum} \label{5ptboundary}
%%%%%%%%%%%%%%%%%%%%

\begin{figure}[!h]
  \centering
    %%%%%%%%%%%%%%%%%%%%
  %%  1
  %%%%%%%%%%%%%%%%%%%%
    \subfigure{
          \begin{picture}(90,80) %(-25,-15)
            \Line(15,35)(15,60)
              \Line(15,35)(0,15)
              \Line(15,35)(30,15)
                    \Line(15,35)(10,15)

                       \BCirc(15,36){8}

                           \Text(15,65)[br]{$n$}
                             \Text(0,45)[br]{$s_{21}$}
                           \Text(0,5)[br]{$1$}
                           \Text(10,5)[br]{$2$}
                             \Text(25,5)[br]{$\dots$}
                           \Text(50,5)[br]{$n-1$}
                         \Text(80,25)[br]{$+\dots$}
\end{picture}
}
  %%%%%%%%%%%%%%%%%%%%
  %%  2
  %%%%%%%%%%%%%%%%%%%%
    \subfigure{
          \begin{picture}(70,80) (-30,0)
            \Line(15,35)(15,60)
              \Line(15,35)(0,15)
              \Line(15,35)(30,15)
                    \Line(15,35)(20,15)

                       \BCirc(15,36){8}

                           \Text(15,65)[br]{$n$}
                             \Text(10,45)[br]{$(s_{21}+\dots +s_{2,n-1})$}
                           \Text(0,5)[br]{$1$}
                           \Text(20,5)[br]{$2$}
                             \Text(15,5)[br]{$\dots$}
                           \Text(50,5)[br]{$n-1$}
                                   \Text(80,25)[br]{$=$}
\end{picture}
}
  %%%%%%%%%%%%%%%%%%%%
  %%  osd
  %%%%%%%%%%%%%%%%%%%%
    \subfigure[]{
      \begin{picture}(130,90) (-35,-15)
    \Line(40,50)(40,35)
    \Line(15,20)(40,35)
    \Line(65,20)(40,35)

%    \Line(15,20)(0,0)
%    \Line(15,20)(15,0)
%    \Line(15,20)(30,0)

    \Line(65,20)(50,0)
%    \Line(65,20)(65,0)
    \Line(65,20)(80,0)

%    \BCirc(15,20){8}
    \BCirc(65,20){8}

            \Text(40,55)[br]{$n$}
%            \Text(0,-10)[br]{$1$}
            \Text(15,10)[br]{$2$}
%            \Text(47,-10)[br]{$n-1$}

                 \Text(50,-10)[br]{$1$}
                 \Text(75,-10)[br]{$\dots$}
                 \Text(105,-10)[br]{$n-1$}
                           \Text(25,40)[br]{$-\,p_{n}^2$}
                                               \Text(120,20)[br]{$.$}
%                                               \Text(-70,20)[br]{$+$}
    \end{picture}\label{osd}
    }
    \caption{The graph representation of off-shell BCJ relation.}\label{offshellgraph}

\end{figure}
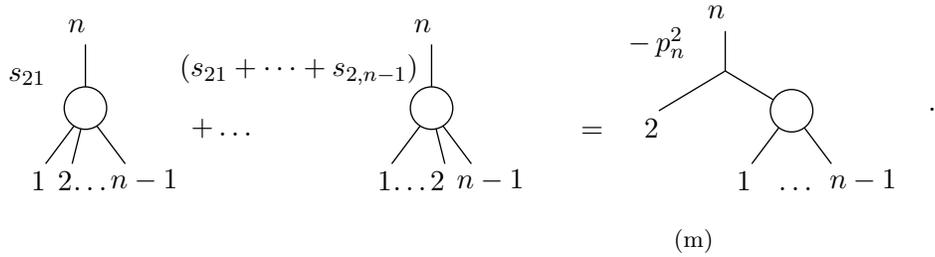

Before showing that the boundary terms in the integral
(\ref{eq:intbcj}) cancel, here we present a generalized scalar BCJ
relation to be used later in the proof, where the leg  $n$  carries
an off-shell momentum (but still color-ordered):

\begin{eqnarray}
&
s_{21}\tilde{A}(123\dots;\, n)+(s_{21}+s_{23})\tilde{A}(132\dots;\, n)
+\dots(s_{21}+\dots s_{2,n-1})\tilde{A}(13\dots n-1,2;\, n)
\nonumber \\
& \hspace{8cm} = - \sum_{color~c}p_{n}^{2}\,
f^{n2c}\frac{1}{P_{c}^{2}}\tilde{A}(1,3\dots n-1;\, c)
\label{eq:offshellbcj}
\end{eqnarray}
It is clear that equation (\ref{eq:offshellbcj}) is identical to the
fundamental  BCJ relation (\ref{BCJ-fund}) when $p_{n}$ is on-shell.
The relation is easily illustrated in terms of Figure
\ref{offshellgraph}

\newpage

%

%\newpage

We prove the off-shell relation (\ref{eq:offshellbcj}) by
induction. At 3-point, equation (\ref{eq:offshellbcj}) simply
restates the algebraic relation between momenta $p_{1}$, $p_{2}$ and
$p_{3}$
\begin{equation}
s_{21}\tilde{A}(12;\,3)=p_{3}^{2}\, \tilde{A}(12;\,3)\end{equation}
where we have used the degenerated case $A(3;c)= \delta_{3}^c P_c^2$
and the antisymmetric property $f^{321}=-f^{123}$.

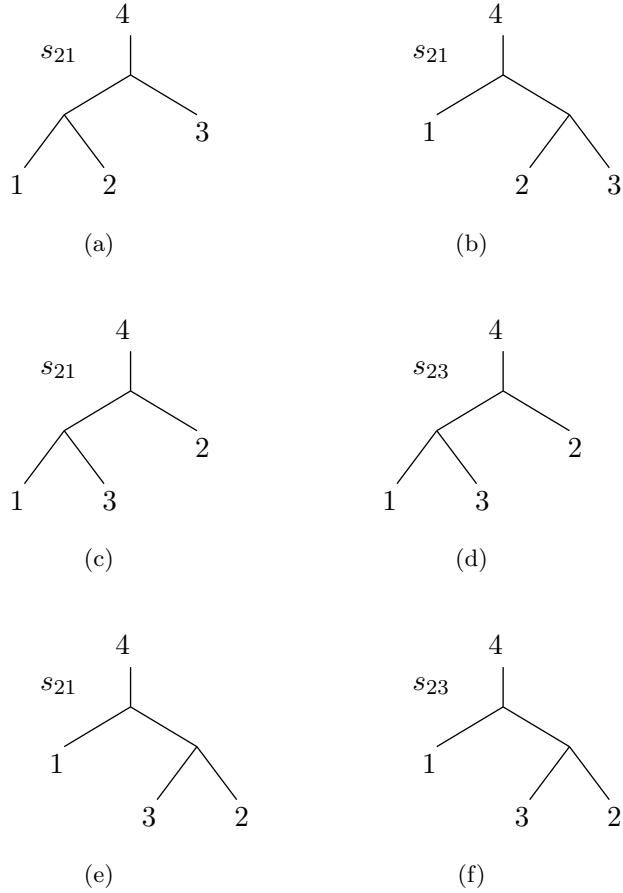
\begin{figure}[!h]
  \centering
  %%%%%%%%%%%%%%%%%%%%
  %%  1
  %%%%%%%%%%%%%%%%%%%%
    \subfigure[]{
      \begin{picture}(130,90) (-35,-15)
    \Line(40,50)(40,35)
    \Line(15,20)(40,35)
    \Line(65,20)(40,35)

    \Line(15,20)(0,0) % left pitchfork left
    \Line(15,20)(30,0) % left pitchfork right

%    \Line(65,20)(50,0) % right pitchfork left
%    \Line(65,20)(80,0) % left pitchfork right
%    \Line(65,20)(65,0)
%    \Line(15,20)(15,0)
%    \BCirc(15,20){8} % left circle
%    \BCirc(65,20){8} % right circle

            \Text(40,55)[br]{$4$}
            \Text(0,-10)[br]{$1$}
%            \Text(15,10)[br]{$1$}
            \Text(35,-10)[br]{$2$}

%                 \Text(50,-10)[br]{$3$}
                 \Text(70,10)[br]{$3$}
                 %\Text(75,-10)[br]{$\dots$}
%                 \Text(85,-10)[br]{$2$}
                           \Text(20,40)[br]{$s_{21}$}
                                             %  \Text(120,20)[br]{$.$}
%                                               \Text(-70,20)[br]{$+$}
    \end{picture}\label{5ptos1}
    }
  %%%%%%%%%%%%%%%%%%%%
  %%  2
  %%%%%%%%%%%%%%%%%%%%
    \subfigure[]{
      \begin{picture}(130,90) (-35,-15)
    \Line(40,50)(40,35)
    \Line(15,20)(40,35)
    \Line(65,20)(40,35)

%    \Line(15,20)(0,0) % left pitchfork left
%    \Line(15,20)(30,0) % left pitchfork right

    \Line(65,20)(50,0) % right pitchfork left
    \Line(65,20)(80,0) % left pitchfork right
%    \Line(65,20)(65,0)
%    \Line(15,20)(15,0)
%    \BCirc(15,20){8} % left circle
%    \BCirc(65,20){8} % right circle

            \Text(40,55)[br]{$4$}
%            \Text(0,-10)[br]{$1$}
            \Text(15,10)[br]{$1$}
%            \Text(47,-10)[br]{$n-1$}

                 \Text(50,-10)[br]{$2$}
                 %\Text(75,-10)[br]{$\dots$}
                 \Text(85,-10)[br]{$3$}
                           \Text(20,40)[br]{$s_{21}$}
                                             %  \Text(120,20)[br]{$.$}
%                                               \Text(-70,20)[br]{$+$}
    \end{picture}\label{5ptos2}
    }
  \\
    %%%%%%%%%%%%%%%%%%%%
  %%  3
  %%%%%%%%%%%%%%%%%%%%
    \subfigure[]{
      \begin{picture}(130,90) (-35,-15)
    \Line(40,50)(40,35)
    \Line(15,20)(40,35)
    \Line(65,20)(40,35)

    \Line(15,20)(0,0) % left pitchfork left
    \Line(15,20)(30,0) % left pitchfork right

%    \Line(65,20)(50,0) % right pitchfork left
%    \Line(65,20)(80,0) % left pitchfork right
%    \Line(65,20)(65,0)
%    \Line(15,20)(15,0)
%    \BCirc(15,20){8} % left circle
%    \BCirc(65,20){8} % right circle

            \Text(40,55)[br]{$4$}
            \Text(0,-10)[br]{$1$}
%            \Text(15,10)[br]{$1$}
            \Text(35,-10)[br]{$3$}

%                 \Text(50,-10)[br]{$3$}
                 \Text(70,10)[br]{$2$}
                 %\Text(75,-10)[br]{$\dots$}
%                 \Text(85,-10)[br]{$2$}
                           \Text(20,40)[br]{$s_{21}$}
                                             %  \Text(120,20)[br]{$.$}
%                                               \Text(-70,20)[br]{$+$}
    \end{picture}\label{5ptos3}
    }
  %%%%%%%%%%%%%%%%%%%%
  %%  4
  %%%%%%%%%%%%%%%%%%%%
    \subfigure[]{
      \begin{picture}(130,90) (-35,-15)
    \Line(40,50)(40,35)
    \Line(15,20)(40,35)
    \Line(65,20)(40,35)

    \Line(15,20)(0,0) % left pitchfork left
    \Line(15,20)(30,0) % left pitchfork right

%    \Line(65,20)(50,0) % right pitchfork left
%    \Line(65,20)(80,0) % left pitchfork right
%    \Line(65,20)(65,0)
%    \Line(15,20)(15,0)
%    \BCirc(15,20){8} % left circle
%    \BCirc(65,20){8} % right circle

            \Text(40,55)[br]{$4$}
            \Text(0,-10)[br]{$1$}
%            \Text(15,10)[br]{$1$}
            \Text(35,-10)[br]{$3$}

%                 \Text(50,-10)[br]{$3$}
                 \Text(70,10)[br]{$2$}
                 %\Text(75,-10)[br]{$\dots$}
%                 \Text(85,-10)[br]{$2$}
                           \Text(20,40)[br]{$s_{23}$}
                                             %  \Text(120,20)[br]{$.$}
%                                               \Text(-70,20)[br]{$+$}
    \end{picture}\label{5ptos4}
    }\\
  %%%%%%%%%%%%%%%%%%%%
  %%  5
  %%%%%%%%%%%%%%%%%%%%
    \subfigure[]{
      \begin{picture}(130,90) (-35,-15)
    \Line(40,50)(40,35)
    \Line(15,20)(40,35)
    \Line(65,20)(40,35)

%    \Line(15,20)(0,0) % left pitchfork left
%    \Line(15,20)(30,0) % left pitchfork right

    \Line(65,20)(50,0) % right pitchfork left
    \Line(65,20)(80,0) % left pitchfork right
%    \Line(65,20)(65,0)
%    \Line(15,20)(15,0)
%    \BCirc(15,20){8} % left circle
%    \BCirc(65,20){8} % right circle

            \Text(40,55)[br]{$4$}
%            \Text(0,-10)[br]{$1$}
            \Text(15,10)[br]{$1$}
%            \Text(47,-10)[br]{$n-1$}

                 \Text(50,-10)[br]{$3$}
                 %\Text(75,-10)[br]{$\dots$}
                 \Text(85,-10)[br]{$2$}
                           \Text(20,40)[br]{$s_{21}$}
                                             %  \Text(120,20)[br]{$.$}
%                                               \Text(-70,20)[br]{$+$}
    \end{picture}\label{5ptos5}
    }
    %%%%%%%%%%%%%%%%%
    %%
    %%  6
    %%%%%%%%%%%%%%%%%
        \subfigure[]{
      \begin{picture}(130,90) (-35,-15)
    \Line(40,50)(40,35)
    \Line(15,20)(40,35)
    \Line(65,20)(40,35)

%    \Line(15,20)(0,0) % left pitchfork left
%    \Line(15,20)(30,0) % left pitchfork right

    \Line(65,20)(50,0) % right pitchfork left
    \Line(65,20)(80,0) % left pitchfork right
%    \Line(65,20)(65,0)
%    \Line(15,20)(15,0)
%    \BCirc(15,20){8} % left circle
%    \BCirc(65,20){8} % right circle

            \Text(40,55)[br]{$4$}
%            \Text(0,-10)[br]{$1$}
            \Text(15,10)[br]{$1$}
%            \Text(47,-10)[br]{$n-1$}

                 \Text(50,-10)[br]{$3$}
                 %\Text(75,-10)[br]{$\dots$}
                 \Text(85,-10)[br]{$2$}
                           \Text(20,40)[br]{$s_{23}$}
                                             %  \Text(120,20)[br]{$.$}
%                                               \Text(-70,20)[br]{$+$}
    \end{picture}\label{5ptos6}
    }
\caption{The boundary contributions of four-point BCJ
relation.}\label{5ptosbcj}

\end{figure}

To get a clearer picture of how a proof is constructed
when generalized to $n$-points,  let us consider
applying equation (\ref{eq:offshellbcj}) to 4-point amplitudes. At
4-points, the left side of the equation is given by
\begin{equation}
s_{21}\tilde{A}(123;\,4)+(s_{21}+s_{23})\tilde{A}(132;\,4),\label{eq:4ptbcjexample}
\end{equation}
which can as well be represented by a summation over the six graphs
from  Figure\ref{5ptosbcj} (\ref{5ptos1} to \ref{5ptos6}). These
figures are obtained by the Feynman diagram expansion multiplied by
the corresponding kinematic factors $s_{21}$ and $s_{23}$.

\newpage

We see that graphs \ref{5ptos2} and \ref{5ptos5} cancel each other
as the result of the antisymmetry of the vertex connecting legs $2$
and $3$. In graph \ref{5ptos1} and graph \ref{5ptos6}, the internal
lines are canceled by kinematic factors $s_{21}$ and $s_{23}$
respectively, thus we can combine them using  Jacobi identity to get
Figure \ref{5ptos7}:

\begin{figure}[!h]
  \centering
  %%%%%%%%%%%%%%%%%%%%
  %%  1
  %%%%%%%%%%%%%%%%%%%%
    \subfigure{
      \begin{picture}(100,90) (0,-15)
    \Line(40,50)(40,35)
    \Line(15,20)(40,35)
    \Line(65,20)(40,35)

    \Line(15,20)(0,0) % left pitchfork left
    \Line(15,20)(30,0) % left pitchfork right

%    \Line(65,20)(50,0) % right pitchfork left
%    \Line(65,20)(80,0) % left pitchfork right
%    \Line(65,20)(65,0)
%    \Line(15,20)(15,0)
%    \BCirc(15,20){8} % left circle
%    \BCirc(65,20){8} % right circle

            \Text(40,55)[br]{$4$}
            \Text(0,-10)[br]{$1$}
%            \Text(15,10)[br]{$1$}
            \Text(35,-10)[br]{$2$}

%                 \Text(50,-10)[br]{$3$}
                 \Text(70,10)[br]{$3$}
                 %\Text(75,-10)[br]{$\dots$}
%                 \Text(85,-10)[br]{$2$}
                           \Text(20,40)[br]{$s_{21}$}
                                             %  \Text(120,20)[br]{$.$}
                                               \Text(100,20)[br]{$+$}
    \end{picture}
    }
    %%%%%%%%%%%%%%%%%
    %%
    %%  6
    %%%%%%%%%%%%%%%%%
        \subfigure{
      \begin{picture}(100,90) (0,-15)
    \Line(40,50)(40,35)
    \Line(15,20)(40,35)
    \Line(65,20)(40,35)

%    \Line(15,20)(0,0) % left pitchfork left
%    \Line(15,20)(30,0) % left pitchfork right

    \Line(65,20)(50,0) % right pitchfork left
    \Line(65,20)(80,0) % left pitchfork right
%    \Line(65,20)(65,0)
%    \Line(15,20)(15,0)
%    \BCirc(15,20){8} % left circle
%    \BCirc(65,20){8} % right circle

            \Text(40,55)[br]{$4$}
%            \Text(0,-10)[br]{$1$}
            \Text(15,10)[br]{$1$}
%            \Text(47,-10)[br]{$n-1$}

                 \Text(50,-10)[br]{$3$}
                 %\Text(75,-10)[br]{$\dots$}
                 \Text(85,-10)[br]{$2$}
                           \Text(20,40)[br]{$s_{23}$}
                                             %  \Text(120,20)[br]{$.$}
%                                               \Text(-70,20)[br]{$+$}
    \end{picture}
    }
  %%%%%%%%%%%%%%%%%%%%
  %%  4
  %%%%%%%%%%%%%%%%%%%%
    \subfigure[]{
      \begin{picture}(100,90) (-20,-15)
    \Line(40,50)(40,35)
    \Line(15,20)(40,35)
    \Line(65,20)(40,35)

    \Line(15,20)(0,0) % left pitchfork left
    \Line(15,20)(30,0) % left pitchfork right

%    \Line(65,20)(50,0) % right pitchfork left
%    \Line(65,20)(80,0) % left pitchfork right
%    \Line(65,20)(65,0)
%    \Line(15,20)(15,0)
%    \BCirc(15,20){8} % left circle
%    \BCirc(65,20){8} % right circle

            \Text(40,55)[br]{$4$}
            \Text(0,-10)[br]{$1$}
%            \Text(15,10)[br]{$1$}
            \Text(35,-10)[br]{$3$}

%                 \Text(50,-10)[br]{$3$}
                 \Text(70,10)[br]{$2$}
                 %\Text(75,-10)[br]{$\dots$}
%                 \Text(85,-10)[br]{$2$}
                           \Text(20,40)[br]{$s_{31}$}
                                             %  \Text(120,20)[br]{$.$}
                                               \Text(-20,20)[br]{$=$}
    \end{picture}\label{5ptos7}
    }
\caption{The combination of (a) and (f) of Figure 2.}

\end{figure}
Since the same tree structure is shared between  \ref{5ptos3},
\ref{5ptos4} and \ref{5ptos7}, the summation of these three graphs
is then given by the sum over the attached kinematic factors
$(s_{21}+s_{23}+s_{31})$, multiplied by the common factors
represented by the tree, yielding
\begin{figure}[!h]
  \centering
  %%%%%%%%%%%%%%%%%%%%
  %%  4
  %%%%%%%%%%%%%%%%%%%%
    \subfigure{
      \begin{picture}(100,90) (-20,-15)
    \Line(40,50)(40,35)
    \Line(15,20)(40,35)
    \Line(65,20)(40,35)

    \Line(15,20)(0,0) % left pitchfork left
    \Line(15,20)(30,0) % left pitchfork right

%    \Line(65,20)(50,0) % right pitchfork left
%    \Line(65,20)(80,0) % left pitchfork right
%    \Line(65,20)(65,0)
%    \Line(15,20)(15,0)
%    \BCirc(15,20){8} % left circle
%    \BCirc(65,20){8} % right circle

            \Text(40,55)[br]{$4$}
            \Text(0,-10)[br]{$1$}
%            \Text(15,10)[br]{$1$}
            \Text(35,-10)[br]{$3$}

%                 \Text(50,-10)[br]{$3$}
                 \Text(70,10)[br]{$2$}
                 %\Text(75,-10)[br]{$\dots$}
%                 \Text(85,-10)[br]{$2$}
                           \Text(20,40)[br]{$(s_{21}+s_{23}+s_{31})$}
                                             %  \Text(120,20)[br]{$.$}
                                           %    \Text(-20,20)[br]{$=$}
    \end{picture}
    }
     %%%%%%%%%%%%%%%%%%%%
  %%  4
  %%%%%%%%%%%%%%%%%%%%
    \subfigure{
      \begin{picture}(100,90) (-20,-15)
    \Line(40,50)(40,35)
    \Line(15,20)(40,35)
    \Line(65,20)(40,35)

%    \Line(15,20)(0,0) % left pitchfork left
%    \Line(15,20)(30,0) % left pitchfork right

    \Line(65,20)(50,0) % right pitchfork left
    \Line(65,20)(80,0) % left pitchfork right
%    \Line(65,20)(65,0)
%    \Line(15,20)(15,0)
%    \BCirc(15,20){8} % left circle
%    \BCirc(65,20){8} % right circle

            \Text(40,55)[br]{$4$}
%            \Text(0,-10)[br]{$1$}
            \Text(15,10)[br]{$2$}
%            \Text(35,-10)[br]{$3$}

                 \Text(50,-10)[br]{$1$}
%                 \Text(70,10)[br]{$2$}
                 %\Text(75,-10)[br]{$\dots$}
                 \Text(85,-10)[br]{$3$}
                           \Text(20,40)[br]{$-p_{4}^{2}$}
                                             %  \Text(120,20)[br]{$.$}
                                               \Text(-20,20)[br]{$=$}
    \end{picture}\label{5ptos8}
    }
\caption{The off-shell BCJ relation of four-point amplitude.}

\end{figure}

from which we arrive at the graphical representation of
$-p_{4}^{2}\, f^{42c}\frac{1}{P_{c}^{2}}\,\tilde{A}(13;c)$ as was
claimed in (\ref{eq:offshellbcj}).

\vspace{1cm}

\textbf{Proving $n$-point off-shell relations by induction}

\begin{figure}[!h]
  \centering
  %%%%%%%%%%%%%%%%%%%%
  %%  osa
  %%%%%%%%%%%%%%%%%%%%
    \subfigure[]{
      \begin{picture}(160,90) (-25,-15)
    \Line(40,50)(40,35)
    \Line(15,20)(40,35)
    \Line(65,20)(40,35)

    \Line(15,20)(0,0)
    \Line(15,20)(15,0)
    \Line(15,20)(30,0)

    \Line(65,20)(50,0)
%    \Line(65,20)(65,0)
    \Line(65,20)(80,0)

    \BCirc(15,20){8}
    \BCirc(65,20){8}

            \Text(40,55)[br]{$n$}
            \Text(0,-10)[br]{$1$}
            \Text(15,-10)[br]{$2$}
            \Text(30,-10)[br]{$i$}

                 \Text(62,-10)[br]{$i+1$}
                 \Text(77,-10)[br]{$\dots$}
                 \Text(105,-10)[br]{$n-1$}

                                  \Text(0,40)[br]{$s_{21}$}
                    \Text(120,20)[br]{$+\dots$}
    \end{picture} \label{osa}
    }
  %%%%%%%%%%%%%%%%%%%%
  %%  osb
  %%%%%%%%%%%%%%%%%%%%
    \subfigure[]{
      \begin{picture}(160,90) (-25,-15)
    \Line(40,50)(40,35)
    \Line(15,20)(40,35)
    \Line(65,20)(40,35)

    \Line(15,20)(0,0)
%    \Line(15,20)(15,0)
    \Line(15,20)(30,0)

    \Line(65,20)(50,0)
    \Line(65,20)(65,0)
    \Line(65,20)(80,0)

    \BCirc(15,20){8}
    \BCirc(65,20){8}

            \Text(40,55)[br]{$n$}
            \Text(0,-10)[br]{$1$}
            \Text(27,-10)[br]{$\dots$}
            \Text(30,-10)[br]{$i$}

                 \Text(47,-10)[br]{$2$}
                 \Text(74,-10)[br]{$i+1$}
                 \Text(105,-10)[br]{$n-1$}
                           \Text(20,40)[br]{$(s_{21}+\dots+s_{2i})$}
                                               \Text(120,20)[br]{$+\dots$}
    \end{picture} \label{osb}
    }
    \\
      %%%%%%%%%%%%%%%%%%%%
  %%  osc
  %%%%%%%%%%%%%%%%%%%%
    \subfigure[]{
      \begin{picture}(130,90) (-55,-15)
    \Line(40,50)(40,35)
    \Line(15,20)(40,35)
    \Line(65,20)(40,35)

    \Line(15,20)(0,0)
%    \Line(15,20)(15,0)
    \Line(15,20)(30,0)

%    \Line(65,20)(50,0)
%    \Line(65,20)(65,0)
%    \Line(65,20)(80,0)

    \BCirc(15,20){8}
%    \BCirc(65,20){8}

            \Text(40,55)[br]{$n$}
            \Text(0,-10)[br]{$1$}
            \Text(23,-10)[br]{$\dots$}
            \Text(47,-10)[br]{$n-1$}

%                 \Text(62,-10)[br]{$i+1$}
                 \Text(75,10)[br]{$2$}
%                 \Text(105,-10)[br]{$n-1$}
                           \Text(20,40)[br]{$(s_{21}+\dots+s_{2,n-1})$}
%                                               \Text(100,20)[br]{$=$}
                                               \Text(-70,20)[br]{$\dots+$}
    \end{picture}\label{osc}
    }
\caption{The Feynman diagrams organized according to the vertex leg
$n$ attached.} \label{n-vertex}

\end{figure}
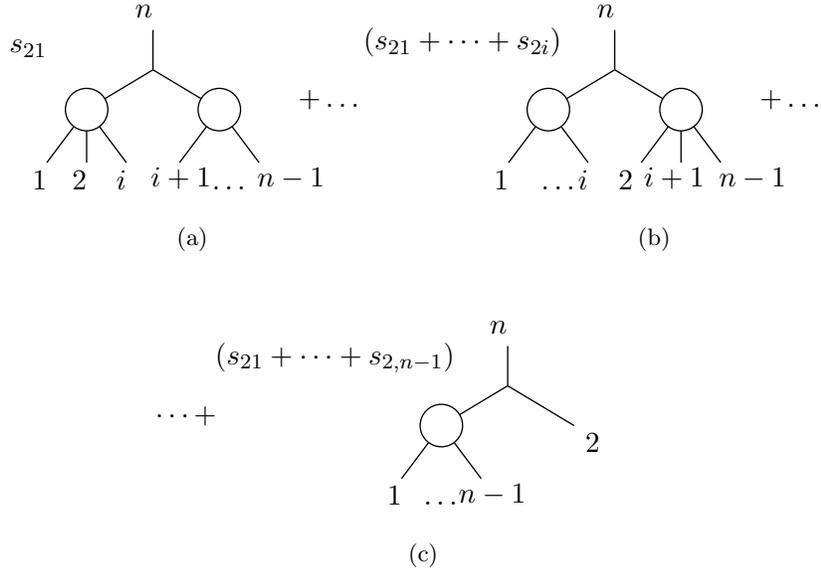

For general $n$-point relation we note that amplitudes on the left
side of the off-shell relation (\ref{eq:offshellbcj}) can be
reorganized according to how the rest of the external legs are
connecting to leg $n$ (see Figure \ref{n-vertex}). For example, for
amplitude $A(1,3,..,i,2,i+1,...,n)$, all Feynman diagrams will be of
the form drawn in Figure (\ref{osa}) where legs $1,3,...,k$
constitue a subtree connecting to leg $n$ through  the left
propagator and $k,...,n-1$ constitue another subtree connecting to
leg $n$ through the right propagator. In these diagrams leg $2$ can
be inserted into either the left subtree or into the right subtree.

%%%%%%%%%%%%%%%%%%%%%%%%%%%%%%%%%%%%%%%%%%%%%%%%%%%%
%%%%%%%%%%%%%%%%%%%%%%%%%%%%%%%%%%%%%%%%%%%%%%%%%%%%
%%%%%%%%%%%%%%%%%%%%%%%%%%%%%%%%%%%%%%%%%%%%%%%%%%%%

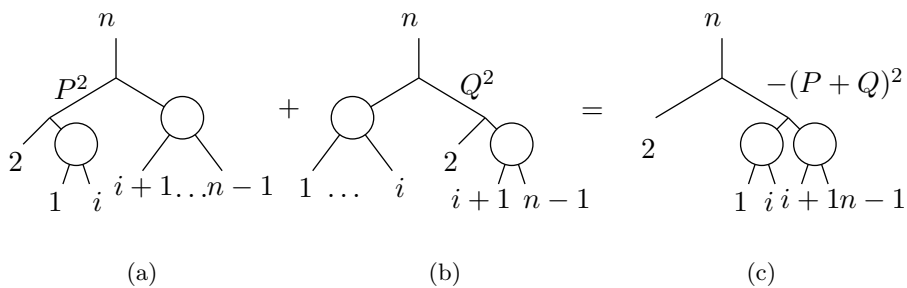
\begin{figure}[!h]
  \centering
  %%%%%%%%%%%%%%%%%%%%
  %%  ose
  %%%%%%%%%%%%%%%%%%%%
    \subfigure[]{
      \begin{picture}(100,80) (0,-25)
    \Line(40,50)(40,35)
    \Line(15,20)(40,35)
    \Line(65,20)(40,35)

%    \Line(15,20)(0,0)
%    \Line(15,20)(15,0)
%    \Line(15,20)(30,0)
        \Line(15,20)(5,10)
        \Line(15,20)(25,10)
                    \Line(25,10)(30,-5)
                    \Line(25,10)(20,-5)
          \BCirc(25,10){8}

    \Line(65,20)(50,0)
%    \Line(65,20)(65,0)
    \Line(65,20)(80,0)

%    \BCirc(15,20){8}
    \BCirc(65,20){8}

            \Text(40,55)[br]{$n$}
            \Text(5,0)[br]{$2$}
            \Text(20,-15)[br]{$1$}
            \Text(35,-15)[br]{$i$}

                 \Text(62,-10)[br]{$i+1$}
                 \Text(77,-10)[br]{$\dots$}
                 \Text(100,-10)[br]{$n-1$}

                                  \Text(30,27)[br]{$P^{2}$}
                    \Text(110,20)[br]{$+$}
    \end{picture} \label{ose}
    }
      %%%%%%%%%%%%%%%%%%%%
  %%  osf
  %%%%%%%%%%%%%%%%%%%%
    \subfigure[]{
      \begin{picture}(100,80) (0,-25)
    \Line(40,50)(40,35)
    \Line(15,20)(40,35)
    \Line(65,20)(40,35)

    \Line(15,20)(0,0)
%    \Line(15,20)(15,0)
    \Line(15,20)(30,0)

        \Line(65,20)(55,10)
        \Line(65,20)(75,10)
                    \Line(75,10)(80,-5)
                    \Line(75,10)(70,-5)
%          \BCirc(25,10){8}

           %    \Line(65,20)(50,0)
%    \Line(65,20)(65,0)
           % \Line(65,20)(80,0)

    \BCirc(15,20){8}
    \BCirc(75,10){8}

            \Text(40,55)[br]{$n$}
            \Text(0,-10)[br]{$1$}
            \Text(20,-10)[br]{$\dots$}
            \Text(35,-10)[br]{$i$}

                 \Text(55,0)[br]{$2$}
                 \Text(75,-15)[br]{$i+1$}
                 \Text(105,-15)[br]{$n-1$}

                                  \Text(69,27)[br]{$Q^{2}$}
                    \Text(110,20)[br]{$=$}
    \end{picture} \label{osf}
    } %%%%%%%%%%%%%%%%%%%%
  %%  osg
  %%%%%%%%%%%%%%%%%%%%
    \subfigure[]{
      \begin{picture}(110,80) (0,-25)
    \Line(40,50)(40,35)
    \Line(15,20)(40,35)
    \Line(65,20)(40,35)

%    \Line(15,20)(0,0)
%    \Line(15,20)(15,0)
%    \Line(15,20)(30,0)

        \Line(65,20)(55,10)
        \Line(65,20)(75,10)
                    \Line(75,10)(80,-5)
                    \Line(75,10)(70,-5)

                    \Line(55,10)(60,-5)
                    \Line(55,10)(50,-5)
%          \BCirc(25,10){8}

           %    \Line(65,20)(50,0)
%    \Line(65,20)(65,0)
           % \Line(65,20)(80,0)

    \BCirc(55,10){8}
    \BCirc(75,10){8}

            \Text(40,55)[br]{$n$}
            \Text(15,5)[br]{$2$}

            \Text(50,-15)[br]{$1$}
            \Text(60,-15)[br]{$i$}

%                 \Text(55,0)[br]{$2$}
                 \Text(85,-15)[br]{$i+1$}
                 \Text(110,-15)[br]{$n-1$}

                                  \Text(109,27)[br]{$-(P+Q)^{2}$}
                   % \Text(120,20)[br]{$=$}
    \end{picture} \label{osg}
    }

\caption{The simplification of the left and right subtrees in Figure
5.}\label{fig-6}

    \end{figure}

Now we consider the type of Feynman diagrams, where legs
$1,3,4,...,i$ are at the left subtree and $i+1,...,n-1$ are at the
right subtree while leg $2$ is inserted freely into all possible
positions (the degenerated case given in Figure (\ref{osc}) is not
included). When leg $2$ is at the left subtree (see Figure
(\ref{osa})), the sum over these left sub-trees with the kinematic
factors is nothing but the form (\ref{eq:offshellbcj}) with less
number of particles, thus by the induction, the result can be
represented by the Figure (\ref{ose}) using the graphical identity
Figure \ref{offshellgraph}.

When leg $2$ is at the right subtrees (the type given by Figure
(\ref{osb})), things will be a little bit complicated. First we divide the
kinematic factor $(s_{21}+...+s_{2i}+s_{2 (i+1)}+...+s_{2k})$
 into two parts: the common factor $(s_{21}+...+s_{2i})$
and the remaining factor $(s_{2 (i+1)}+...+s_{2k})$  when leg $2$ is
right behind  leg $k$. For the common factor $(s_{21}+...+s_{2i})$,
when we sum up all possible insertions of leg $2$ at the right
subtree, we get zero by using the $U(1)$-decoupling equation (as we
have remarked, it is true even with one leg off-shell). For the
remaining factor $(s_{2 (i+1)}+...+s_{2k})$, the sum is again in the
form of (\ref{eq:offshellbcj}), then by the induction we obtain the
Figure (\ref{osf}) using the Figure \ref{offshellgraph}.

After the substitutions we are left with Figures (\ref{osc}),
(\ref{ose}) and (\ref{osf}) multiplied by kinematic factor
$\sum_{j=1}^{n-1} s_{2j}$, $(p_1+p_3+..+p_i)^2\equiv P^2$ and
$(p_{i+1}+ p_{i+2}+...+p_{n-1})^2 \equiv Q^2$ respectively. If we
effectively regard the subtree containing legs $1,3,...,i$ as a new
leg carrying momentum $P$ and the subtree containing legs
$i+1,...,n-1$ as another new leg carrying momentum $Q$,  Figure
(\ref{ose}) with factor $P^2$ and  Figure (\ref{osf}) with factor
$Q^2$ can be combined to give Figure (\ref{osg}) with factor
$-(P+Q)^2$ using  Jacobi identity. Thus we see clearly why the
Jacobi identity is crucial for the BCJ relation to be true. As a
last step we combine Figure (\ref{osc}) and (\ref{osg}). Using the
anti-symmetric property, algebraically combining kinematic factors
$-(P+Q)^2+ 2k_2\cdot k_n=- p_n^2$, and then summing  over  $i$ to
obtain an amplitude $\tilde{A}(1,3,...,n-1;c)$, we arrive at exactly
the same result as was claimed in (\ref{eq:offshellbcj}).

In fact, the proof of (\ref{eq:offshellbcj}) reduces to a proof of
fundamental BCJ relation given in (\ref{BCJ-fund}) when we take
$p_n$ on-shell. In other words, we have given a diagrammatic proof
of BCJ relation. We shall see in the following section that the
cancelation of boundary terms in our original
 proof of BCJ relation via  BCFW recursion also requires the
identity (\ref{eq:offshellbcj}).

%%%%%%%%%%%%%%%%%%%%%%%
\subsection{Boundary terms}
%%%%%%%%%%%%%%%%%%%%%%

As we have shown, for the color-ordered scalar theory, there are
boundary contributions from the contour  $\oint {dz\over
z}I_{n}(z)$. These  boundary contributions come from two places when
we write $s_{21}(z)=s_{21}+z s_{q1}$: the first place is boundary
contribution from amplitudes with $(1,n)$ nearby and the second
place is $z s_{q1}$ with ${1\over z}$ part of amplitudes.

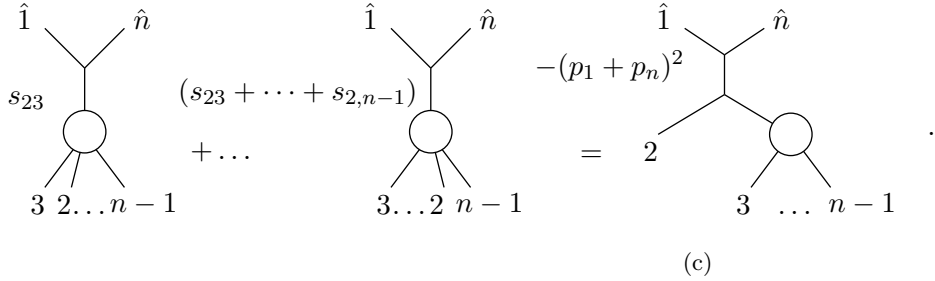
\begin{figure}[!h]
  \centering
    %%%%%%%%%%%%%%%%%%%%
  %%  bca
  %%%%%%%%%%%%%%%%%%%%
    \subfigure{
          \begin{picture}(90,80) %(-25,-15)
            \Line(15,35)(15,60)
              \Line(15,35)(0,15)
              \Line(15,35)(30,15)
                    \Line(15,35)(10,15)

                       \BCirc(15,36){8}

              \Line(15,60)(0,75)
              \Line(15,60)(30,75)
             \Text(-5,75)[br]{$\hat{1}$}
            \Text(40,75)[br]{$\hat{n}$}
%                           \Text(15,65)[br]{$n$}
                             \Text(0,45)[br]{$s_{23}$}
                           \Text(0,5)[br]{$3$}
                           \Text(10,5)[br]{$2$}
                             \Text(25,5)[br]{$\dots$}
                           \Text(50,5)[br]{$n-1$}
                         \Text(80,25)[br]{$+\dots$}
\end{picture}
}
  %%%%%%%%%%%%%%%%%%%%
  %%  bca
  %%%%%%%%%%%%%%%%%%%%
    \subfigure{
          \begin{picture}(70,80) (-30,0)
            \Line(15,35)(15,60)
              \Line(15,35)(0,15)
              \Line(15,35)(30,15)
                    \Line(15,35)(20,15)

                       \BCirc(15,36){8}

              \Line(15,60)(0,75)
              \Line(15,60)(30,75)
             \Text(-5,75)[br]{$\hat{1}$}
            \Text(40,75)[br]{$\hat{n}$}
%                           \Text(15,65)[br]{$n$}
                             \Text(10,45)[br]{$(s_{23}+\dots +s_{2,n-1})$}
                           \Text(0,5)[br]{$3$}
                           \Text(20,5)[br]{$2$}
                             \Text(15,5)[br]{$\dots$}
                           \Text(50,5)[br]{$n-1$}
                                   \Text(80,25)[br]{$=$}
\end{picture}
}
    \subfigure[]{
      \begin{picture}(130,90) (-35,-15)

    \Line(40,50)(25,60)
    \Line(40,50)(55,60)

    \Line(40,50)(40,35)
    \Line(15,20)(40,35)
    \Line(65,20)(40,35)

%    \Line(15,20)(0,0)
%    \Line(15,20)(15,0)
%    \Line(15,20)(30,0)

    \Line(65,20)(50,0)
%    \Line(65,20)(65,0)
    \Line(65,20)(80,0)

%    \BCirc(15,20){8}
    \BCirc(65,20){8}
            \Text(20,60)[br]{$\hat{1}$}
            \Text(65,60)[br]{$\hat{n}$}
%            \Text(0,-10)[br]{$1$}
            \Text(15,10)[br]{$2$}
%            \Text(47,-10)[br]{$n-1$}

                 \Text(50,-10)[br]{$3$}
                 \Text(75,-10)[br]{$\dots$}
                 \Text(105,-10)[br]{$n-1$}
                           \Text(25,40)[br]{$-(p_{1}+p_{n})^2$}
                                               \Text(120,20)[br]{$.$}
%                                               \Text(-70,20)[br]{$+$}
    \end{picture} \label{bca}
    }
\caption{The boundary contribution of first source.} \label{fig7}
\end{figure}

Contributions from the first place is given by
$A(n,1,-P_{n1}){1\over P_{n1}^2} $ multiplying with the following factor
\bean s_{21}A(P_{n1}, 2,3,...,n-1)+ (s_{21}+s_{23})
A(P_{n1},3,2,4,...,n-1)\ +...+ (s_{21}+\sum_{j=3}^{n-1} s_{j2}
A(P_{n1},3,4,...,n-1,2)~. \eean
In the equation above, terms carrying the same factor $s_{21}$  sum up to give zero from
$U(1)$-decoupling identity, while the remaining  terms assume the form of
(\ref{eq:offshellbcj}), which we replace by
\bea -\sum_{a,c} A(n,1,-P_{n1}^a) f^{a2c} A(3,4,...,n-1;
P_{n12}^c)~, \eea
Graphically this is represented by Figure \ref{fig7} (\ref{bca}).

%can arise either from a $z$ independent kinematic factor $S_{2I}$
%multiplied by a boundary term in the amplitude or from a shifted
%$s_{2\hat{1}}(z)$ multiplied by a term in the amplitude that behaves
%as $1/z$ asymptotically. The first case corresponds to graphs where
%both shifted legs $\hat{1}$ and $\hat{n}$ are attached to the same
%vertex. From off-shell relation these graphs sum up and give

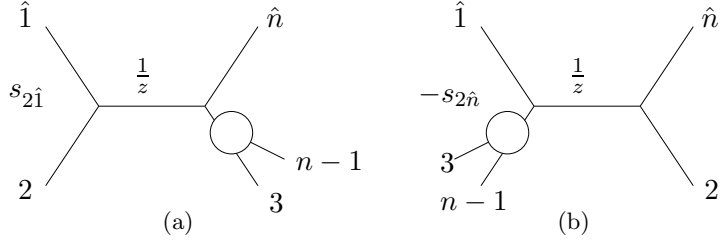
\begin{figure}[!h]
  \centering
  %%%%%%%%%%%%%%%%%%%%
  %%  bcb
  %%%%%%%%%%%%%%%%%%%%
    \subfigure[]{
% \fcolorbox{white}{white}{
  \begin{picture}(140,70) (-20,0)
    \SetWidth{0.375}

\Line(20,30)(0,0)
%\Line(20,30)(0,30)
\Line(20,30)(0,60)

\Line(20,30)(60,30)

\Line(60,30)(80,0)
%\Line(60,30)(80,30)
\Line(60,30)(80,60)
      \Line(70,20)(90,10)
%    \BCirc(20,30){10}
    \BCirc(70,20){8}

        \Text(-5,60)[br]{$\hat{1}$}
%        \Text(-5,30)[br]{$2$}
        \Text(-5,-5)[br]{$2$}
                \Text(90,-10)[br]{$3$}
                \Text(120,5)[br]{$n-1$}
        \Text(0,30)[br]{$s_{2\hat{1}}$}
        \Text(40,35)[br]{$\frac{1}{z}$}
        \Text(90,60)[br]{$\hat{n}$}
  \end{picture} \label{bcb}
}
%%%%%%%%%%%%%%%%%%%%%%%%%%
%%   bcc
%%%%%%%%%%%%%%%%%%%%%%%%%%
    \subfigure[]{
% \fcolorbox{white}{white}{
  \begin{picture}(130,70) (-30,0)
    \SetWidth{0.375}

\Line(20,30)(0,0)
%\Line(20,30)(0,30)
\Line(20,30)(0,60)

\Line(20,30)(60,30)

\Line(60,30)(80,0)
%\Line(60,30)(80,30)
\Line(60,30)(80,60)
      \Line(10,20)(-10,10)
%    \BCirc(20,30){10}
    \BCirc(10,20){8}

        \Text(-5,60)[br]{$\hat{1}$}
        \Text(-10,5)[br]{$3$}
        \Text(10,-10)[br]{$n-1$}
%                \Text(90,-10)[br]{$3$}
%                \Text(120,5)[br]{$n-1$}
        \Text(0,30)[br]{$-s_{2\hat{n}}$}
        \Text(90,-5)[br]{$2$}
        \Text(40,35)[br]{$\frac{1}{z}$}
        \Text(90,60)[br]{$\hat{n}$}
  \end{picture} \label{bcc}
}

\caption{The two degenerated of boundary contributions from second
source.}\label{fig8}
\end{figure}

Contributions from the second place come from diagrams that have only one
 propagator between the shifted legs $\hat{1}$ and $\hat{n}$,
such as the one illustrated in Figure \ref{fig8} (\ref{bcb}). More
explicitly, the left cubit vertex has  leg $1$, propagator with left
subtree containing
 legs $3,...,i$, and the propagator connecting $1,n$, while the
right cubic vertex has  leg $n$, propagator with left subtree
containing some legs $i+1,...,2$, and the propagator connecting
$1,n$. Thus except the two degenerated cases given by Figure (\ref{bcb})
and (\ref{bcc}), the general expression will have the form
\bea \left\{A(1, -P_L, -P_L-p_1) {1\over P_L^2}
A_L(P_L,3,...,i)\right\}\times \left\{ A(P_L+p_1, -P_R, n) {1\over
P_R^2} A_R(P_R,i+1,...,n-1)\right\}\eea
where leg $2$ can be inserted at any allowed position in $A_L$ and
$A_R$. It is easy to see that the sum over all insertions of leg $2$
in $A_L$ and $A_R$ gives zero by $U(1)$-decoupling identity.

From the above analysis, we see that sum of  boundary contributions
reduces to three terms given by Figure (\ref{bca}), (\ref{bcb}) and
(\ref{bcc}). If we take the subtree made by leg $3,4,..,n-1$ as a
new leg $P_{3(n-1)}$, then the sum of the three diagrams are nothing
but the three terms that appear in Jacobi identity and we have a
complete cancelation.

%%%%%%%%%%%%%%%%%%%%%%%%%%%%%%%%%%%%%%%%%%%%%%%%%%%%%
%%%%%%%%%%%%%%%%%%%%%%%%%%%%%%%%%%%%%%%%%%%%%%%%%%%%%
%%%%%%%%%%%%%%%%%%%%%%%%%%%%%%%%%%%%%%%%%%%%%%%%%%%%%
%%%
%%%
%%%    KLT
%%%
%%%%%%%%%%%%%%%%%%%%%%%%%%%%%%%%%%%%%%%%%%%%%%%%%%%%%
%%%%%%%%%%%%%%%%%%%%%%%%%%%%%%%%%%%%%%%%%%%%%%%%%%%%%
%%%%%%%%%%%%%%%%%%%%%%%%%%%%%%%%%%%%%%%%%%%%%%%%%%%%%

%%%%%%%%%%%%%%%%%%%%%%%%%%%%
\section{Field theory proof of color KLT relations}\label{sec:KLT}
%%%%%%%%%%%%%%%%%%%%%%%%%%%

After the  long preparation above, finally in this section we present
a pure field theory proof of the KLT relation proposed in
\cite{Bern:1999bx}. Having shown that the KK-relation and BCJ
relation are also true for color-ordered scalar theory, the proof of
the proposed KLT relation will be much similar to the one for the
KLT relation between gravity and gauge theory in
\cite{BjerrumBohr:2010ta, BjerrumBohr:2010zb,
BjerrumBohr:2010yc,Feng:2010br}, where more details can be found.

%%%%%%%%%%%%%%%%%%
\subsection{The KLT relation}
%%%%%%%%%%%%%%%%%%

First we need to present the KLT relations. To do so we introduce
following  function ${\cal S}$ given by
\bea {\cal S}[i_1,...,i_k|j_1,j_2,...,j_k]_{p_1} & = & \prod_{t=1}^k
(s_{i_t 1}+\sum_{q>t}^k \theta(i_t,i_q) s_{i_t
i_q})~,~~~\label{S-def}\eea
where $\theta(i_t, i_q)$ is zero when  pair $(i_t,i_q)$ has the same
ordering at both set ${\cal I},{\cal J}$ and otherwise, it is one.
%%The second definition is of a  dual function${\cal \W S}$ given by
%
%%\bea {\cal {\W
%%S}}[i_2,..,i_{n-1}|j_2,...,j_{n-1}]_{p_n}=\prod_{t=2}^{n-1} (s_{j_t
%%n}+ \sum_{q<t} \theta(j_t, j_q) s_{j_t j_q})~.~~\label{dual-S-def}
%%\eea
%
%where $\theta$ has same definition as in ${\cal S}$.
With above
definitions, the KLT relation given originally in \cite{Bern:1998sv}
can be written as
\bea M_n & = & (-)^{n+1}\sum_{\sigma\in S_{n-3}}\sum_{\a\in
S_j}\sum_{\b\in S_{n-3-j}}
A(1,\{\sigma_2,..,\sigma_j\},\{\sigma_{j+1},..,\sigma_{n-2}\},
n-1,n) {\cal S}[
\a(\sigma_2,..,\sigma_j)|\sigma_2,..,\sigma_j]_{p_1}\nn & & \times
{\cal S}[\sigma_{j+1},..,\sigma_{n-2}
|\b(\sigma_{j+1},..,\sigma_{n-2})]_{p_{n-1}} \W
A(\a(\sigma_2,..,\sigma_j),1,n-1,\b(\sigma_{j+1},..,\sigma_{n-2}),n)~~~\label{KLT-bern}\eea
where $j=[n/2-1]$ is a fixed number determined by $n$ and the two
subscripts $p_1, p_{n-1}$ of two ${\cal S}$-functions are also
important to distinguish. However, using the BCJ relation we can
reduce or increase $j$ and we reach following two more symmetric
formula
\bea M_n  =  (-)^{n+1}\sum_{\sigma,\W\sigma\in S_{n-3}}
A(1,\sigma(2,n-2), n-1,n) {\cal S}[
\W\sigma(2,n-2))|\sigma(2,n-2))]_{p_1} \W
A(n-1,n,\W\sigma(2,n-2),1)~~~~~\label{KLT-Pure-S}\eea
as well as
\bea M_n  =  (-)^{n+1}\sum_{\sigma,\W\sigma\in S_{n-3}}
A(1,\sigma(2,n-2), n-1,n) {\cal S}[
\sigma(2,n-2))|\W\sigma(2,n-2))]_{p_{n-1}} \W
A(1,n-1,\W\sigma(2,n-2),n)~~~~~\label{KLT-Dual-S}\eea
Furthermore, a manifestly $(n-2)!$ symmetric KLT formula was
found in \cite{BjerrumBohr:2010ta}
\bea M_n=(-)^n\sum_{\gamma,\b}{\W A(n,\gamma(2,...,n-1),1)  {\cal
S}[ \gamma(2,...,n-1)|\b(2,..,n-1)]_{p_1} A(1,\b(2,...,n-1),n)\over
s_{123..(n-1)}}~~~\label{newKLT}\eea
as well as  its dual form
\bea M_n & = & (-)^n\sum_{\b,\gamma} {A (1,\b(2,...,n-1),n) {\cal
S}[\b(2,...,n-1)|\gamma(2,..,n-1)]_{p_{n}} \W A(n,\gamma(2,...,n-1),
1)\over s_{2...n}}~~~\label{New-KLT-dual} \eea
In \cite{Feng:2010hd}, it was shown explicitly how to regularize
the above expression and the equivalence between the new $(n-2)!$ form
(\ref{newKLT}) and old $(n-3)!$ form (\ref{KLT-Pure-S}) was
proved.

One important property of function ${\cal S}$ is
\bea 0 = I \equiv \sum_{\a\in S_k}{\cal
S}[\a(i_1,...,i_k)|j_1,j_2,...,j_k]_{p_1}
A(k+2,\a(i_1,...,i_k),1)~.~~~\label{I-com}\eea
or the one obtained by color order reversing
\bea 0 =I \equiv \sum_{\a\in S_k}{\cal S}[j_k,...,j_1|\a]_{p_1} A(1,
\a,k+2)~.~~~\label{I-com-2}\eea
%

%%%%%%%%%%%%%%%%%%%%%%%
\subsection{The proof of KLT relation for gauge theory}
%%%%%%%%%%%%%%%%%%%%%%%%

Now we prove that when $A$ is color-ordered tree level
amplitude of gauge theory and $\tilde{A}$ is the color-ordered
scalar theory discussed in previous sections, the $M_n$ is nothing
but the total scattering amplitude ${\cal A}_n$ of gauge theory as
conjectured in \cite{Bern:1999bx}. The idea of our proof will be
the following. With BCFW-deformation of $(1,n)$ we have
\bea B_1 & = & \oint {dz\over z} {\cal A}_n (z)= {\cal
A}_n(z=0)-\sum_\a
{\cal A}_L {1\over P_\a^2} {\cal A}_R, ~~~\label{full-A}\\
 B_2 & = & \oint {dz\over z} M_n(z)= M_n(z=0)+
 \sum_{poles} {\rm Res} \left( {M_n(z)\over z}\right), ~~~\label{eq:BCFWM} \eea
If $B_1=0=B_2$ and  we can use the induction to show that $-\sum_\a
{\cal A}_L {1\over P_\a^2} {\cal A}_R=\sum_{poles} {\rm Res} \left(
{M_n(z)\over z}\right)$, then we have shown ${\cal A}_n= M_n$.

The vanishing of $B_1$ is obvious for gauge theory. For $B_2$, let
us check the formula (\ref{KLT-Dual-S}). First the kinematic factor
${\cal \W S}$ is independent of $z$. Secondly, for nearby $1,n$, the
gauge part contributes a ${1\over z}$ and the scalar part, a ${1\over
z^0}$, thus the overall behavior will be ${1\over z}$ and we get
vanishing boundary contribution $B_2$.

Having established the vanishing of boundary contributions $B_1,
B_2$, we will focus on the pole parts.
%% The proof of color KLT relations follows
%%directly from that of graviton/gluon
%% KLT relations \cite{KLT}. We start by taking BCFW shiftings
%%on legs $1$ and $n-1$ in the expression
%%
%%\begin{equation}
%%M_{n}=(-1)^{n+1}\sum_{\sigma,\tilde{\sigma}\in
%%S_{n-3}}\tilde{A}_{n}(n-1,n,\tilde{\sigma}_{2,n-2},1)\mathcal{S}
%%[\tilde{\sigma}|\sigma]_{p_{1}}A_{n}(1,\sigma_{2,n-2},n-1,n),\label{eq:cKLT1}\end{equation}
%%
%%
%%where $\tilde{A}_{n}$ and $A_{n}$ stand for scalar and gauge field
%%amplitudes respectively. Despite generically scalar amplitudes admit
%%boundary terms as can be shown easily from power counting, the
%%shifted function $M_{n}(z)$ vanishes at large $z$, which is clear
%%from an alternative expression of (\ref{eq:cKLT1}) derived in
%%\cite{KLT},
%%
%%\begin{equation}
%%M_{n}=(-1)^{n+1}\sum_{\sigma,\tilde{\sigma}\in
%%S_{n-3}}A_{n}(1,\sigma_{2,n-2},n-1,n)\tilde{\mathcal{S}}
%%[\sigma_{2,n-2}|\tilde{\sigma}_{2,n-2}]_{p_{n-1}}\tilde{A}_{n}
%%(1,n-1,\tilde{\sigma}_{2,n-2},n)\label{eq:cKLT2}\end{equation}
%%
%%
%%In the dual formula \eqref{eq:cKLT2} an unshifted leg $p_{n-1}$ is
%%adopted instead of $p_{1}$ as the reference momentum in
%%$\tilde{\mathcal{S}}$. While scalar amplitude $\tilde{A}_{n}$ and
%%$\tilde{\mathcal{S}}$ are asymptotically finite the function
%%$M_{n}(z)$ is suppressed by gluon amplitude at large $z$. Without
%%oundary contributions, the usual BCFW applies and we have,
%%
%%
%%which allows the unshifted function $M_{n}(0)$ to be re-expressed in
%%terms of cut amplitudes $A_{L}\frac{1}{P^{2}}A_{R}$.
As in \cite{BjerrumBohr:2010ta, BjerrumBohr:2010zb,
BjerrumBohr:2010yc,Feng:2010br}, we will show inductively that the
BCFW-expansion of $M_{n}$ gives precisely the BCFW-expansion of full
gluon amplitudes. The starting point, i.e., 3-point KLT relation, is
easy to check since the scalar amplitude is simply a structure
constant
\begin{equation}
M_{3}(1^{a},2^{b},3^{c})=f^{abc}A_{3}(1,2,3).\end{equation}

For general $M_{n}$, by the full symmetry,  we can  consider a
representative  cut $P_{123...k}$ with $k=2,...,n-2$ and other cuts
will be easily related by permutations. For this cut, pole structure
in $M_{n}$ can be further
 divided into following three categories as in
\cite{BjerrumBohr:2010ta, BjerrumBohr:2010zb,
BjerrumBohr:2010yc,Feng:2010br}:
\begin{itemize}
\item (i) The pole only shows up in scalar amplitude $\tilde{A}_{n}$.

\item (ii) The pole only shows up in gluon amplitude $A_{n}$.

\item (iii) Both $\tilde{A}_{n}$ and $A_{n}$ contain pole $1/s_{12\dots k}$.
\end{itemize}
Residues in (i) are given by cutting the scalar amplitude and
multiplying $\tilde{\mathcal{S}}$ and $A_{n}$ with legs $1$ and $n$
shifted.
\begin{equation}
\sum_{\sigma,\tilde{\sigma},\alpha}\frac{\tilde{A}(n-1,\hat{n},\tilde{\sigma}_{k+1,n-2},\hat{P})\,\tilde{A}(-\hat{P},\alpha_{2,k},\hat{1})}{s_{12\dots
k}}\mathcal{S}[\tilde{\sigma}_{k+1,n-2}\alpha_{2,k}|\sigma_{2,n-2}]_{\hat{p}_{1}}A_{n}(\hat{1},\sigma_{2,n-2},n-1,\hat{n})\label{eq:casei}\end{equation}
With this configuration, a factor of $\mathcal{S}[\alpha_{2,k}|\rho_{2,k}]_{\hat{p}_1}$
can be extracted from the function
$\mathcal{S}[\tilde{\sigma}_{k+1,n-2}\alpha_{2,k}|\sigma_{2,n-2}]_{\hat{p}_{1}}$
. Using (\ref{I-com}) we find
\begin{equation}
\sum_{\alpha}\tilde{A}(-\hat{P},\alpha_{2,k},\hat{1})\mathcal{S}[\alpha_{2,k}|\rho_{2,k}]=0~.\end{equation}
 Residues in (ii) vanish for
the same reason. In the case (iii) the special ordering of
$A,\tilde{A}$ allows $\mathcal{S}$ to factorize as the following
\begin{equation}
\mathcal{S}[\tilde{\sigma}_{k+1,n-2}\alpha_{2,k}|\beta_{2,k}\sigma_{k+1,n-2}]_{\hat{p}_{1}}=\mathcal{S}[\alpha_{2,k}|\beta_{2,k}]\times\mathcal{S}[\tilde{\sigma}_{k+1,n-2}|\sigma_{k+1,n-2}]_{\hat{P}},\end{equation}
thus the residue can be written as
\begin{eqnarray}
& \frac{1}{s_{12\dots
k}}\sum_{h,color}\sum_{\alpha,\beta}\frac{\tilde{A}(-\hat{P},\alpha_{2,k},\hat{1})\mathcal{S}[\alpha_{2,k}|\beta_{2,k}]_{\hat{p}_{1}}A(\hat{1},\beta_{2,k},-\hat{P}^{h})}{s_{\hat{1}2\dots
k}} \hspace{5cm}
\nonumber \\
&
\times\sum_{\sigma,\tilde{\sigma}}\tilde{A}(n-1,\hat{n},\tilde{\sigma}_{k+1,n-2},\hat{P})\mathcal{S}[\tilde{\sigma}_{k+1,n-2}|\sigma_{k+1,n-2}]_{\hat{P}}A(-\hat{P}^{-h},\sigma_{k+1,n-2},n-1,\hat{n}).\label{eq:caseiii}\end{eqnarray}
The residue is nothing but the  products of sub-amplitudes
$M_{k+1}$ and $M_{n-k+1}$ using Eq. (\ref{KLT-Pure-S}) to Eq.
(\ref{New-KLT-dual})
%, Equation (\ref{eq:caseiii}) reads
%
\begin{equation}
\sum_{h}\frac{M_{k+1}(\hat{1},2,\dots,k,-\hat{P}^{h})\,
M_{n-k+1}(k+1,\dots,\hat{P}^{-h})}{s_{12\dots k}}.\end{equation}
Thus we have shown that the pole part of (\ref{eq:BCFWM}) does match
up with the pole part of (\ref{full-A})by induction and we
have completed the proof \footnote{There is a technical point in our
proof. For the cut without $p_{n-1}$ it is better to use the formula
Eq. (\ref{KLT-Pure-S}), while for cut having $p_{n-1}$ it is better
to use the formula Eq. (\ref{KLT-Dual-S}).}.

%% For poles $1/s_{12\dots n-1\dots k}$ where leg $n-1$ is attached
%%to the subamplitude on the left we repeat the same argument to split
%%the residue as products of two fewer-point functions $M_{k+1}$ and
%%$M_{n-k+1}$. By induction the function $M_{n}$ then equals the color
%%dependent Yang-Mills amplitude.

%%%%%%%%%%%%%%%%%%%%%%%%
\subsection{Another proof by off-shell BCJ relations }
%\label{sec:KLT} %%%%%%%%%%%%%%%%%%%%%%%

\begin{figure}[!h]
  \centering
    %%%%%%%%%%%%%%%%%%%%
  %%  1
  %%%%%%%%%%%%%%%%%%%%
    \subfigure{
          \begin{picture}(90,80) %(-25,-15)
            \Line(15,35)(15,60)
              \Line(15,35)(0,15)
              \Line(15,35)(30,15)
  %                  \Line(15,35)(10,15)

                       \BCirc(15,36){8}

               \Text(-90,20)[br]{$\sum_{S_{n-2}}$}
                           \Text(15,65)[br]{$n$}
                             \Text(0,45)[br]{$\mathcal{S}[2\dots n-1|\{ i \}_{2,n-1}]$}
                           \Text(0,5)[br]{$1$}
                           %\Text(10,5)[br]{$2$}
                             \Text(47,3)[br]{$\{ i \}_{2,n-1}$}
                           %\Text(50,5)[br]{$n-1$}
                         \Text(70,25)[br]{$=$}
\end{picture}
}
  %%%%%%%%%%%%%%%%%%%%
  %%  first step
  %%%%%%%%%%%%%%%%%%%%
    \subfigure{
      \begin{picture}(130,90) (-100,0)
    \Line(40,50)(40,35)
    \Line(15,20)(40,35)
    \Line(65,20)(40,35)

%    \Line(15,20)(0,0)
%    \Line(15,20)(15,0)
%    \Line(15,20)(30,0)

    \Line(65,20)(50,0)
%    \Line(65,20)(65,0)
    \Line(65,20)(80,0)

%    \BCirc(15,20){8}
    \BCirc(65,20){8}

            \Text(40,55)[br]{$n$}
%            \Text(0,-10)[br]{$1$}
            \Text(15,10)[br]{$2$}
%            \Text(47,-10)[br]{$n-1$}

\Text(-90,20)[br]{$\sum_{S_{n-2}}$}
\Text(0,45)[br]{$\mathcal{S}[3\dots n-1|\{ i \}_{3,n-1}]$}
                 \Text(50,-10)[br]{$1$}
                 \Text(95,-12)[br]{$\{ i \}_{3,n-1}$}
              %   \Text(105,-10)[br]{$n-1$}
                           \Text(35,32)[br]{$\left(-\right)\,p_{n}^2$}
                                               %\Text(120,20)[br]{$.$}
%                                               \Text(-70,20)[br]{$+$}
    \end{picture}
    }\\
      %%%%%%%%%%%%%%%%%%%%
  %%  first step
  %%%%%%%%%%%%%%%%%%%%
    \subfigure{
      \begin{picture}(100,90) (0,-20)

       \Line(10,30)(130,30)
       \Line(30,30)(30,0)
       \Line(50,30)(50,0)

       \Line(110,30)(110,0)
       \Text(-30,30)[br]{$= \hspace{0.5cm} \dots \hspace{0.5cm} =$}
       \Text(5,30)[br]{$n$}
       \Text(40,40)[br]{$\left(-\right)^{n}\,p_{n}^{2}$}
       \Text(30,-10)[br]{$2$}
       \Text(50,-10)[br]{$3$}
       \Text(80,0)[br]{$\dots$}
       \Text(130,-10)[br]{$n-1$}
       \Text(140,30)[br]{$1$}

    \end{picture}
    }
    \caption{Graphs obtained by repeated substitutions using off-shell fundamental BCJ relations}
    \label{Off-KLT}
\end{figure}
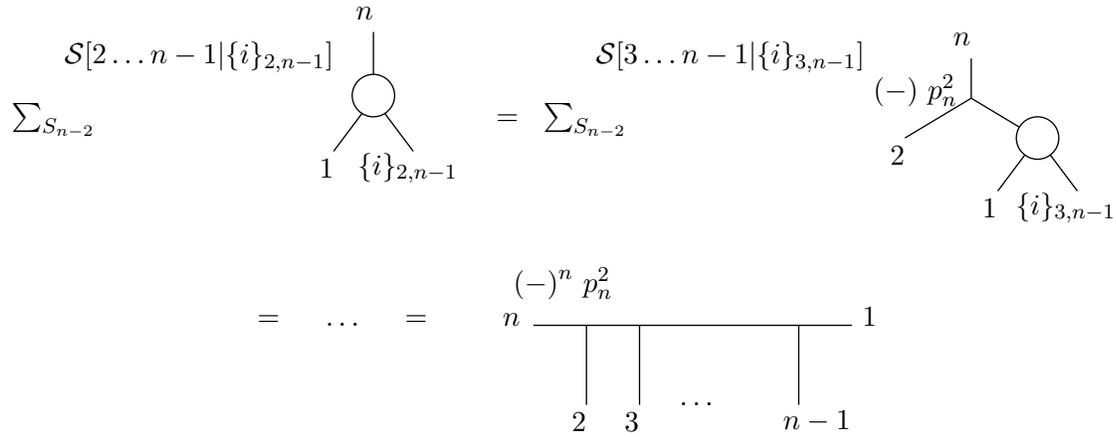

For the gauge KLT relation, there is another direct proof using the
generalized fundamental BCJ relations
discussed in section \ref{5ptboundary}. To start with,  let us consider
following analogous  sum appearing  in the standard BCJ relations
\begin{equation}
\sum_{\{i\}\in S_{n-2}}\mathcal{S}[2,3,\dots n-1|i_{2},i_{3},\dots i_{n-1}]
\,\tilde{A}_{n}(1,\, i_{2},i_{3},\dots i_{n-1};n)\label{eq:offshellbcjx}\end{equation}
 where
as in the previous section we use $\tilde{A}_{n}(1\dots;n)$ to
denote an n-point scalar amplitude  and a semicolon is used to
emphasize that leg $n$ is allowed to be off-shell. Using the
definition of function $\mathcal{S}$ we see that the sum in
(\ref{eq:offshellbcjx}) can be written as
\begin{eqnarray}
& \sum_{\{j\}\in S_{n-3}}\mathcal{S}[3,\dots n-1|j_{3},\dots
j_{n-1}]
\hspace{5cm} \nonumber \\
&
\hspace{0.5cm}\times\left[s_{21}A_{n}(1,2,j_{3}\dots j_{n-1};n)+(s_{21}+s_{2j_{3}})\tilde{A}_{n}(1,j_{3},2,\dots j_{n-1};n)+\dots\right]
\nonumber \\
& =-{p_{n}^{2}\over p_{n2}^2}\, f^{n2c}\sum_{\{i\}\in
S_{n-3}}\mathcal{S}[3,\dots n-1|j_{3},\dots
j_{n-1}]\tilde{A}_{n-1}(1,j_{3},\dots;c).
\end{eqnarray}
where $\{ j \}$ is the set defined by deleting leg $2$ from the set $\{ i \}$ and in the
last line we have used (\ref{eq:offshellbcj}). The sum over ${\cal
S}$ can be done similarly and we obtain
\bea (-)^2{p_{n}^{2}\over p_{n2}^2}\, f^{n2c}{p_{2n}^{2}\over
p_{n23}^2}\, f^{c3c_1}\sum_{\{i\}\in S_{n-4}}\mathcal{S}[4,\dots
n-1|j_{3},\dots j_{n-1}]\tilde{A}_{n-1}(1,j_{4},\dots;c_1) \eea
Repeatedly reducing  the number of legs contained in the amplitude
one by one,  we finally arrive at
\begin{eqnarray}
& \sum_{\{i\}\in S_{n-3}}\mathcal{S}[2,3,\dots
n-1|\beta_{2,n-1}]\,\tilde{A}_{n}(1,\beta_{2,n-1};n)
=(-)^{n-2}p_{n}^{2}\, f^{n2c}f^{c3e}\dots f^{e,(n-1),1},
\label{eq:ofshellbcjz}\end{eqnarray}
The result (\ref{eq:ofshellbcjz}) can be represented graphic as
Figure \ref{Off-KLT}.

%\vspace{1cm}

Now the proof of gluon KLT relations can be obtained directly from
application of the off-shell relations (\ref{eq:ofshellbcjz}) to
following $(n-2)!$ symmetric form
\begin{equation}
M_{n}=(-1)^{n}\sum_{\gamma,\beta}\frac{{A}_{n}(n,\gamma_{2,n-1},1)
\mathcal{S}[\gamma_{2,n-1}|\beta_{2,n-1}]_{p_{1}}\tilde{A}_{n}(1,\beta_{2,n-1},n)}
{s_{12\dots,n-1}},\label{eq:KLT1}\end{equation}
 and we arrive at the familiar
expression discovered by Del Duca, Dixon and
Maltoni\cite{DelDuca:1999rs} which expresses a gluon total amplitude
as products of structure constants multiplied by color-ordered
amplitudes
\begin{equation}
M_{n}=\sum_{\gamma}f^{n,i_{2},c}f^{c,i_{3},e} \dots
f^{e,i_{n-1},1}A_{n}(n,i_{2},\dots i_{n-1},1).\end{equation}
%

%%%%%%%%%%%%%%%%%%%%%%%%%%%%%%%%%%%%%%
%%%%%%%%%%%%%%%%%%%%%%%%%%%%%%%%%%%%%%

%%%%%%%%%%%%%%%%%%
\section{Conclusion}\label{sec:con}
%%%%%%%%%%%%%%%%%%%%
In this paper, we systematically study the color-ordered scalar
theory and prove the color-order reversed relation,
$U(1)$-decoupling identity, KK and BCJ relations for this field. In
our proof, the BCFW on-shell recursion relation with nontrivial
boundary contributions plays the central role. Having established
the KK and BCJ relations, we prove the gauge KLT conjecture made in
\cite{Bern:1999bx}, where the full gauge amplitude can be factorized
as the product of color-ordered gauge amplitude and the
color-ordered scalar amplitude. A  byproduct in our proof is the
generalized fundamental BCJ relation with one leg off-shell and
using it, both BCJ relation and KLT relation can be derived
directly.

% We then use the BCJ relation to prove the color-KLT relation in which any gluon
% are factorized into a vector and a scalar with color.
% In the proof of KK relation, we introduce the generalized $U(1)$-decoupling
%  identity. We also extend this generalized $U(1)$-decoupling identity
% to gluon amplitudes and explain its physical meaning.

There are a few things worth to notice. As we have mentioned, in the
proof of BCJ relation, the generalized fundamental BCJ relation with
one leg off-shell is founded and using it (\ref{eq:ofshellbcjz})
gives the direct relation between color and ${\cal S} \tilde{A}$
which is not obvious at all. Recently there is a nice suggestion
about the dual form of gauge theory given in \cite{Bern:2011ia}. If
we can find some generalized fundamental BCJ relation with one leg
off-shell in the gauge theory, then we maybe possible to give a good
construction of dual color factor $n_i$ as discussed in
\cite{Bern:2011ia}.

Another interesting thing is  generalized KLT relations conjectured
in \cite{Bern:1999bx} besides the one studied here, such as the tree
amplitudes with gluons coupled to gravitons can be factorized into
pure-gluon color-ordered amplitudes and amplitudes with scalar
coupled to gluons, or be factorized into fermion pair coupled to
gluons. KLT relation has also be found for those amplitudes with a
massive graviton exchange when we factorize external gluons into two
quarks. Though these KLT relations are more complicated than the
pure gluon case, as stated in \cite{Bern:1999bx}, it is possible to
study them using similar idea as given in this paper.

%%%%%%%%%%%%%%%%%%%%%%
\subsection*{Acknowledgements}
%%%%%%%%%%%%%%%%%%%%%%

We  would like to thank discussions with P. Vanhove, T. Sondergaard
and the hospitality of Kavli Institute for Theoretical Physics,
China (KITPC) and Kavli Institute for Theoretical Physics, Santa
Barbara (KITP), USA, where part of this work was done. We are
supported by fund from Qiu-Shi, the Fundamental Research Funds for
the Central Universities with contract number 2010QNA3015, the
Chinese NSF funding under contract No.10875104, No.11031005, as well
as the USA National Science Foundation under Grant No. NSF
PHY05-51164.

%%%%%%%%%%%%%%%%%%%%%%
\appendix
%%%%%%%%%%%%%%%%%%%%%%

%%%%%%%%%%%%%%
\section{The generalized $U(1)$-decoupling identity}
%%%%%%%%%%%%%%
In the proof of KK-relation, there is one new identity we need to
use to show the vanishing of boundary contributions. It is given by
\bea \sum_{P(O(\a)\bigcup O(\b))} A(1, P(O(\a)\bigcup O(\b)))=0
~~~\label{new-identity}\eea
where the sum is over all permutations where relative ordering of
two subsets $\a,\b$ is kept. This identity has been shown to be the
sum of two KK-relations, so it is true if the KK-relation is true.
However, relation (\ref{new-identity}) has a physical picture as the
generalized $U(1)$-decoupling identity as we will show in this
section.
For simplicity we will focus on the pure gauge theory with $U(N)$
gauge group. The generator can be chosen as $N\times N$ matrix
$E_{ij}$ with number $1$ at position $(i,j)$ and zero at other
places. These $N^2$ generators can be divided into following three
categories: category (A) with $1\leq i,j\leq N_1$; category (B) with
$N_1+1\leq i,j\leq N$ and category (C) with $1\leq i\leq N_1,
N_1+1\leq j\leq N$ or $1\leq j\leq N_1, N_1+1\leq i\leq N$. The Lie
bracket is $[E_{ij}, E_{kl}]=\delta_{jk} E_{il}-\delta_{il} E_{kj}$,
so we have
\bea [A,A]\sim A,~~[B,B]\sim B,~~[C,C]\sim A+B,~~[A,B]\sim
0,~~[A,C]\sim C,~~[B,C]\sim C~.~~\label{Lie-braked}\eea
The trace is given by $Tr(E_{ij} E_{kl})=\delta_{jk} \delta_{il}$ so
we have
\bea tr(A A)\neq 0,~~tr(BB)\neq 0,~~~tr(CC)\neq
0,~~tr(AB)=0,~~tr(AC)=0,~~~tr(BC)=0~.~~\label{UN-trace}\eea
The trace structure (\ref{UN-trace}) tells us that the propagator
can only be the $AA, BB, CC$ three types and there is no mixing
between different categories. This observation is very important for
our late arguments. Having (\ref{Lie-braked}) and (\ref{UN-trace}),
from the Lagrangian
\bea Tr( (\partial A+[A,A]) (\partial A+[A,A]))~.~~\label{YM-action}
\eea
we see that nonzero cubic vertex will be following types
\bea AAA, BBB,  ACC,   BCC, ~~\label{cubic-UN} \eea
and the nonzero four-point vertex will be following types
\bea  AAAA, BBBB, CCCC,  AA CC, BB CC, AB
CC~.~~\label{4point-UN}\eea
Now we consider the amplitude with $1$ belonging to type (C), set
$\a$ belonging to type (A) and set $\b$ belonging to type (B). Under
the color decomposition, the contribution to this particular color
configuration is nothing, but the combination at the left hand side
of (\ref{new-identity}). To show the amplitude is zero for above
color configuration, let us study Feynman diagrams of tree-level
amplitudes. Since the coexistence of type (A) and (B), there must be
propagators of $CC$ type, thus  we see that each diagram must have a
bone structure given by propagators of type $CC$. Since each nonzero
vertex with (C)-type must have at least two legs with type (C) from
(\ref{cubic-UN}) and (\ref{4point-UN}) and the bone structure given
by propagators of type $CC$ must have at least two endpoints, there
is no diagram for color configuration with only one external
particle belonging to type (C). Thus the amplitude of this
particular color configuration must be zero. If there are at least
two external particles belonging to type (C), there are nonzero
contributions from Feynman diagrams and we can not claim anything.
Thus we have given a physical picture of (\ref{new-identity}) by our
generalized $U(1)$-decoupling argument.

\end{document}